\documentclass[sigconf]{acmart}

\usepackage{makecell}
\usepackage{subcaption}
\usepackage{amsthm}
\usepackage{graphicx}
\usepackage{nth}
\usepackage[ruled]{algorithm2e}
\usepackage{color}
\usepackage{gensymb}
\usepackage{multirow}
\newtheorem{defn}{Definition}
\usepackage[font=small]{caption}
\usepackage{dsfont}
\usepackage{wrapfig}

\usepackage{tikz}
\usetikzlibrary{backgrounds}
\usetikzlibrary{bayesnet}
\usetikzlibrary{shapes,decorations,arrows,calc,arrows.meta,fit,positioning}
\tikzset{
    -Latex,auto,node distance =1 cm and 1 cm,semithick,
    state/.style ={ellipse, draw, minimum width = 0.6 cm},
    point/.style = {circle, draw, inner sep=0.04cm,fill,node contents={}},
    bidirected/.style={Latex-Latex,dashed},
    el/.style = {inner sep=2pt, align=left, sloped}
}

\usepackage{pifont}%
\newcommand{\cmark}{\ding{51}}%

\let\oldquote\quote
\let\endoldquote\endquote
\renewenvironment{quote}[2][]
  {\if\relax\detokenize{#1}\relax
     \def\quoteauthor{#2}%
   \else
     \def\quoteauthor{#2~---~#1}%
   \fi
   \oldquote}
  {\par\nobreak\smallskip\hfill(\quoteauthor)%
   \endoldquote\addvspace{\bigskipamount}}

\AtBeginDocument{%
  }

\copyrightyear{2023} 
\acmYear{2023} 
\setcopyright{rightsretained} 
\acmConference[FAccT '23]{2023 ACM Conference on Fairness, Accountability, and Transparency}{June 12--15, 2023}{Chicago, IL, USA}
\acmBooktitle{2023 ACM Conference on Fairness, Accountability, and Transparency (FAccT '23), June 12--15, 2023, Chicago, IL, USA}
\acmDOI{10.1145/3593013.3594042}
\acmISBN{979-8-4007-0192-4/23/06}

\acmSubmissionID{331}

\begin{document}

\title{Gender Animus Can Still Exist Under Favorable Disparate Impact: 
a Cautionary Tale from Online P2P Lending
}

\author{Xudong Shen}
\email{xudong.shen@u.nus.edu}
\orcid{0000-0001-9549-0986}
\affiliation{%
  \institution{National University of Singapore}
  \country{Singapore}
}

\author{Tianhui Tan}
\email{tant@nus.edu.sg}
\orcid{0000-0002-8038-7887}
\affiliation{%
  \institution{National University of Singapore}
  \country{Singapore}
}

\author{Tuan Q. Phan}
\email{tphan@hku.hk}
\orcid{0000-0002-6512-8158}
\affiliation{%
  \institution{The University of Hong Kong}
  \country{Hong Kong, China}
}

\author{Jussi Keppo}
\email{keppo@nus.edu.sg}
\orcid{0000-0002-4571-5566}
\affiliation{%
  \institution{National University of Singapore}
  \country{Singapore}
}

\renewcommand{\shortauthors}{Shen et al.}

\begin{abstract}

This paper investigates gender discrimination and its underlying drivers on a prominent Chinese online peer-to-peer (P2P) lending platform.
While existing studies on P2P lending focus on disparate treatment (DT), DT narrowly recognizes direct discrimination and overlooks indirect and proxy discrimination, providing an incomplete picture.
In this work, we measure a broadened discrimination notion called disparate impact (DI), which encompasses any disparity in the loan's funding rate that does not commensurate with the actual return rate.
We develop a two-stage predictor substitution approach to estimate DI from observational data.
Our findings reveal (\textit{i}) female borrowers, given identical actual return rates, are 3.97\% more likely to receive funding, (\textit{ii}) at least $37.1\%$ of this DI favoring female is indirect or proxy discrimination, and (\textit{iii}) DT indeed underestimates the overall female favoritism by $44.6\%$.
However, we also identify the overall female favoritism can be explained by one specific discrimination driver, rational \emph{statistical discrimination}, wherein investors accurately predict the \emph{expected return rate} from imperfect observations.
Furthermore, female borrowers still require 2\% higher expected return rate to secure funding, indicating another driver taste-based discrimination co-exists and is \emph{against female}.
These results altogether tell a cautionary tale: on one hand, P2P lending provides a valuable alternative credit market where the affirmative action to support female naturally emerges from the rational crowd; on the other hand, while the overall discrimination effect (both in terms of DI or DT) favors female, concerning taste-based discrimination can persist and can be obscured by other co-existing discrimination drivers, such as statistical discrimination.

\end{abstract}



\received{20 February 2007}
\received[revised]{12 March 2009}
\received[accepted]{5 June 2009}

\maketitle

\section{Introduction}
Online peer-to-peer (P2P) lending is a fast-growing FinTech innovation which allows individual investors to directly crowdfund individual borrowers' loans through online platforms, bypassing traditional intermediaries like banks.
In this digital lending landscape, the P2P lending platforms often leverages machine learning (ML) models and alternative data to assess borrower creditworthiness. 
The ML-based credit scoring largely supports investors' lending decisions and improves borrowers' access to credit\cite{larrimore2011peer,emekter2015evaluating}.
As a result, P2P lending has showed great success in extending financial access and promoting financial inclusion~\cite{tang2019peer,demirgucc2022global}.
Notably, the global P2P lending market was valued at USD 82.3 billion in 2021, and is projected to reach USD 804.2 billion by 2030~\cite{p2p_lending_market_forecast}.

The extent of gender discrimination against borrowers on P2P lending platforms, whether male or female, is an important yet under-explored topic.  
While some studies show that female borrowers are often equally~\cite{barasinska2014crowdfunding,chen2020gender} or more likely~\cite{pope2011s,ravina2019love,chen2017gender} to be funded on their P2P loans, they focus on \textit{disparate treatment} (DT), a notion that narrowly recognizes direct discrimination.
Specifically, the existing studies test the direct effect of borrower gender on the loan's funding success, controlling for all other observed characteristics.
However, DT fails to account for the effects of \emph{indirect} and \emph{proxy discrimination}~\cite{Washington_v_Davis}, wherein seemingly neutral practices can still have a disproportionately negative impact on a specific group.
For instance, in \textit{Griggs v. Duke Power Co.}~\cite{Griggs_v_Duke} the US Supreme Court deemed the company's standardized testing requirement for job transfers to be illegal. It is because although the requirement was not intentionally discriminatory, it had a disparate impact on black employees and was not reasonably job-related.
Thus, to fully realize the legal guarantees of equal citizenship, it is crucial to adopt a broader perspective beyond DT.

More importantly, the extent of discrimination elimination will depend on the discrimination drivers~\cite{cooter1994market,fang2011theories}, but identifying discrimination drivers, such as taste-based discrimination~\cite{becker1971economics} and statistical discrimination~\cite{phelps1972statistical,arrow1973theory,aigner1977statistical}, is a non-trivial task.
In taste-based discrimination\footnote{We will use gender taste, gender animus, and taste-based discrimination interchangeably.}, the decision makers discriminate because they have personal taste that favors (or disfavors) members of a particular group.
On the other hand, statistical discrimination refers to the situation where the decision makers perceive the group-level difference in decision-related characteristics and use this information to infer the \emph{expectations} from noisy decision signals.
Unlike taste-based discrimination, statistical discrimination is considered rational and economical.
However, little is known about whether taste-based and statistical discrimination can \emph{co-exist} and contribute to the overall discrimination in online P2P lending.

Online P2P lending provides a unique opportunity to study gender discrimination and its underlying drivers in ML-assisted collective human decisions. 
In this study, we introduce a broadened discrimination notion called disparate impact (DI), which is motivated from the legal doctrine of disparate impact~\cite{Griggs_v_Duke}.
Disparate impact allows the plaintiff to initiate a discrimination case by showing a disparate adverse impact on the protected group \emph{without showing explicit categorization}.
To refute the claim, the defendant must prove the shown disparity can be ``justified as serving a legitimate business goal''~\cite{EECO_v_Metal}.
In other words, DI encompasses any unwarranted disparity that does not commensurate with individuals' true qualifiedness, regardless of whether it is direct discrimination. 
In online P2P lending, since investors can decide the amount they wish to invest in each loan, the loan's eligibility for funding is solely determined by its \textit{return rate}\footnote{We define the return rate of a loan as the proportion of the borrower's actual repayment to the loan's principal. For example, a return rate of 0 signifies a complete loss of both the principal and interest. 
Meanwhile, a return rate of 1 implies that the borrower only repays the principal but not the interest.
We note that default might still occur when return rate $>1$, but default will not occur when the return rate equals ($1 + \text{the interest rate}$).
}.
As such, we define DI as the disparate funding of loans that yield identical return rates but differ in the gender of the borrower.
The notion of DI studied in this work relates to and expands upon the concept of equality of opportunity in computer science~\cite{hardt2016equality,berk2021fairness,xudong2022pami} and disparate impact in economics~\cite{ayres2005three,ayres2010testing,arnold2020measuring}.

However, empirical estimation of DI remains a challenge. This is mainly because the return rates of unfunded loans are unobservable. 
In a related work, \citet{arnold2020measuring} develop a quasi-experimental solution to measure the broadly defined discrimination in the context of bail decisions.
But this approach has limitations, as (\textit{i}) it only pertains to binary qualifiedness, and (\textit{ii}) it relies on random assignment of decision-makers.
It thus may not be applicable in many other settings, including online P2P lending.
Alternative estimation strategies are needed.

We develop a two-stage predictor substitution (2SPS) approach to estimate DI using only observational data.
In the first stage, we predict and impute the return rates for the unsuccessful loans.
Specifically, we consider the decomposition of return rate as the product between the repayment ratio\footnote{We define the repayment ratio of a loan as the proportion of the total amount owed by the borrower, including both principal and interest, that has been successfully repaid. The repayment ratio takes values in $[0,1]$, with a value of 1 indicating that the loan has been fully repaid and no default has occurred.} and (1 + interest rate), where the interest rate is fully observed.
We construct a survival model~\cite{machin2006survival,cleves2008introduction}, which explicitly models the amount of time it takes for a borrower to default, to predict the repayment ratio of the loan.
In the second stage, we estimate DI non-parametrically.
Taken together, we bootstrap the framework for 500 times to obtain confidence intervals that account for both stages.
We also analyze the asymptotic properties of the 2SPS approach and show its robustness.

We obtain anonymized backend data from one of the largest online P2P lending platforms in China and empirically investigate the gender discrimination following the 2SPS approach.
We find female borrowers, given identical return rates, are 3.97\% (95\% CI: 3.98\%$\sim$3.95\%) more likely to be funded.
While the DI favoring female is marginal for loans with return rate $\leq 1.16$, it is highly significant for loans with return rate $>1.16$, which constitute $~90\%$ of our data.
Through a decomposition analysis, we show that at least 37.1\% of the DI favoring female is due to indirect or proxy discrimination.
Corroboratively, DT indeed underestimates the favoritism towards female borrowers by $44.6\%$, equivalent to a $2.15\%$ higher likelihood of funding for female borrowers compared to male borrowers, \textit{ceteris paribus}.
Our results remain consistent across a number of robustness checks.

We further investigate a decision model which decomposes DI into two components, capturing taste-based and statistical discrimination respectively, by incorporating a threshold test~\cite{simoiu2017problem,pierson2018fast} in the second stage of 2SPS approach.
Surprisingly, we find the overall female favoritism can be fully explained as rational statistical discrimination, which posits that the investors accurately predict the \emph{expected return rate} from noisy repayment ratio (intuitively borrower creditworthiness) signals.
The statistical discrimination favors female because female borrowers are factually less likely to default and have higher repayment ratio.
Consequently, given identical but noisy repayment ratio signals, the \emph{expected return rate} is higher for female borrowers than male borrowers. 
However, despite this, female borrowers still need a higher \emph{expected return rate} of 1.099, compared to 1.079 for male borrowers to secure funding.
This suggests another driver taste-based discrimination co-exists and is against female borrowers.

We conclude this work with a discussion of our empirical results.
We contend that gender gap in financial access still exists~\cite{demirgucc2022global} and females are still disadvantaged in many other credit markets~\cite{wellalage2021bank,chaudhuri2020gender,brock2021discriminatory}.
P2P lending, therefore, complements traditional bank lending by providing an alternative credit market where the affirmative action to support females is driven by the market forces rather than bureaucratic rules, is win-win, and can emerge naturally from the rational crowd.
On the other hand, it is important to recognize that different discrimination drivers can co-exist and contribute to the overall discrimination effect.
This paper illustrates a real-world scenario where while the overall discrimination effect favors female (both in terms of DI and DT), gender animus against female can still persist, because it may be obscured due to a substantial amount of statistical discrimination.
We hope the insights in our study act as a starting point for future research to deepen the current understanding of discrimination drivers in human and/or machine decision decisions.

\section{Related Work}
\noindent\textbf{P2P Lending} \quad
Uncovering the importance of FinTech enabled financial inclusion, there is an increasing number of research on online P2P lending.
One stream of research tries to identify various determinants of funding success and borrower default~\cite{herzenstein2008democratization,iyer2016screening}, including appearance~\cite{duarte2012trust,ravina2019love}, social ties~\cite{lin2013judging,freedman2017information}, and textual descriptions~\cite{larrimore2011peer}.
\citet{herzenstein2011strategic,zhang2012rational} study the online investors' herding behavior.
\citet{berger2009emergence} look into the role of group leaders in crowdfunding.
\citet{lin2016home} find that home bias, the tendency that transactions are more likely to occur when two parties are geographically closer, still exists in online P2P lending.
And \citet{tang2019peer} finds P2P lending is a substitute for traditional bank lending in terms of infra-marginal borrowers, but serves as a complement in terms of small loans.

\noindent\textbf{Gender Discrimination in P2P Lending} \quad
Although female is often discriminated against in traditional bank lending~\cite{alesina2013women,blanchflower2003discrimination,muravyev2009entrepreneurs}, findings from P2P lending tend to show otherwise.
For example, \citet{barasinska2014crowdfunding} find no effect of gender on funding success on a German P2P lending platform.
\citet{pope2011s,ravina2019love} find females are favorably treated on an American P2P lending platform \texttt{Prosper.com}.
Notably, \citet*{chen2017gender} and \citet*{chen2020gender} study P2P lending platforms in China.
While \citet*{chen2020gender} find no significant gender effect on funding success,
\citet*{chen2017gender} find females are more likely to be funded than males.
This line of literature, however, focuses on disparate treatment and provides an incomplete picture of gender discrimination.
They also do not identify how taste-based discrimination and statistical discrimination can co-exist.

\noindent\textbf{Disparate Impact} \quad
In economics, the ideal of recognizing any unwarranted disparity that does not commensurate with individuals' true qualifiedness traces back to \citet{aigner1977statistical}.
\citet{ayres2005three,ayres2010testing} discusses the distinction between testing disparate treatment and disparate impact.
\citet{arnold2020measuring} develop a quasi-experimental tool to measure disparate impact in bail decisions.
\citet{bohren2022systemic} further decompose disparate impact into direct discrimination and systemic discrimination.
The latter is the combined effect of indirect and proxy discrimination.
A closely related notion is equality of opportunity (EOpp) in the computer science literature~\cite{hardt2016equality,berk2021fairness,xudong2022pami}.
Motivated from a binary classification setting, EOpp argues the false negative rate, intuitively the denied opportunity to those who deserve it, should be equalized across different protected groups.
Notably, EOpp similarly narrowly controls for the true qualifiedness.

\noindent\textbf{Taste-based and Statistical Discrimination} \quad
The classical taste-based discrimination is first theorized in~\citet{becker1971economics}, which considers discrimination as blatantly applying disparate decision thresholds.
Then, \citet{phelps1972statistical,arrow1973theory,aigner1977statistical} formalize the statistical discrimination model, where the decision makers do not have an intrinsic preference for one group but are incentivized to use the protected attributes to accurately assess the individual's quality of interest, which might lead to unfair outcomes.
There is an increasing interest to distinguish taste-based and statistical discrimination, for example in healthcare~\cite{balsa2001statistical}, bail decisions~\cite{arnold2020measuring} and roads policing~\cite{marx2022absolute}, as well as algorithmic decisions~\cite{patty2022algorithmic} and reinforcement learning~\cite{duenez2021statistical}.

\section{Background}
\subsection{The P2P Lending Platform}
We collaborate with one of the largest online P2P lending platforms\footnote{We are unable to reveal the name of the platform due to confidentiality and agreement with the platform.} in China (hereinafter referred to as “the platform”).
The platform connects individual investors to individual borrowers across the country and offers unsecured personal loans.
When borrowers submit a loan application, they self-specify the desired loan amount, interest rate, and duration of the installment payments from a range allowed on the platform.
The borrowers are also required to provide a set of loan and borrower information.
Then, a loan listing containing these information is generated on the platform.
The loans are open for investors to view and subscribe for a fixed period of time.
The investors may subscribe a small amount for every loans to diversify the risk.
The platform operates an All-or-Nothing crowdfunding policy.
A loan is considered successful if and only if the borrowing amount is fully reached.
Otherwise, the loan is unsuccessful and the borrower will not receive any money.
For the successful loans, the borrowers make repayments in equated monthly installments (EMI) according to the loan term.

The platform develops a machine learning (ML)-based credit scoring model to support investors' lending decisions, making the loan's funding success a ML-assisted collective human decision.
At every loan application, the credit scoring model categorizes the borrower into credit grades from I to VIII, with grade I / VIII representing the lowest / highest risk.
Then, the credit grade is shown on the loan's information page to assist the investors' lending decisions.
However, our data does not include the credit grade information.

\subsection{Data}
We obtained anonymized backend data on loans and borrowers on the platform.
Our data consists of 1,006,161 loan listings with 12-month installment plans between January 1, 2016 and June 30, 2016.
As listed in Table~\ref{tab:descriptive_table} in Appendix, for every loan, our data includes (\textit{i}) \textbf{borrower characteristics} such as gender, marriage, age, employment, education, whether he/she is a repeated borrower, number of past failed borrowings, number of past aborted borrowings, number of past ontime payments, and number of past late payments; and (\textit{ii}) \textbf{loan characteristics} including borrowing amount, interest rate, whether the loan is requested on the mobile application, and whether the loan is an express loan.
Furthermore, our data also includes whether a loan is successfully funded and, if so, the borrower's installment payment record over the loan term.
In consistent with the platform's operational definition, an installment is considered defaulted if the payment is late for more than 90 days.
Table~\ref{tab:descriptive_table} in Appendix reports the descriptive statistics.

We note that the 2SPS estimate of DI only requires us to include covariates that are predictive of the repayment ratio, but \emph{does not requires us to control all factors that enter the investor's lending decisions}, a fundamental difference from DT.
Two comments are in order.
First, our data in fact includes more attributes, such as the borrower's city, district, and registration time.
But we do not include them because we find them unpredictive through a backward feature selection procedure using the Akaike Information Criterion~\cite{akaike1974new}.
Second, the 2SPS estimate of DI might be biased due to omitted variables in predicting the repayment ratio.
Concrete examples are the loan's credit grade and description, which are not available in our data but are reasonably believed to contain additional information about the borrower's default risk.
But this issue is generally unavoidable.
We discuss this bias theoretically in Sec.~\ref{sec:analysis_of_bias} and empirically test it in Sec.~\ref{sec:analysis_DI}.

We conduct the following pre-processing steps.
First, we drop the loans whose payments are non-conventional for ease of analysis, including (1) those who have some installment partially payed (N=2,806) and (2) those who first default some installment(s) but then pay again (N=3,301).
Second, we winsorize the loans whose amount, age, \# past ontime payments, \# past late payments, \# past failed borrowings, and \# past aborted borrowings fall in the top or bottom 0.5\% quantile to eliminate outliers (N=29,985).
Lastly, we drop loans whose interest rates are lower than $16\%$ (N=293,596), which are always unfunded.
It reflects that the investors on the platform simply do not consider loans whose interest rates are lower than $16\%$.
The final sample consists of 676,473 loans.

\noindent\textbf{Borrower Gender} \quad
Given the centrality of gender discrimination in this paper, it is worth elaborating how the gender attribute is obtained.
When borrowers first sign up on the platform, they are required to provide the 18-digit Chinese Citizen Identity Number.
Then, the borrower's biological sex is extracted from the second last digit: odd numbers are issued to males and even numbers to females.
Therefore, in this paper we treat gender as the binary biological sex.
We acknowledge this treatment risks marginalizing and miscategorizing transgender and non-binary gender.
Finally, we note that borrower gender is explicitly shown on the loan's information page.

\subsection{Notations}
\label{sec:notations}
The loans are indexed using a subscript $i\in[I]$.
We use $\mathbf{X}_i$ to denote the vector of loan and borrower characteristics shown in Table~\ref{tab:descriptive_table}.
$G_i\in\{m,f\}$ denotes the borrower's gender: $m$ means male and $f$ means female.
$R_i>0$ denotes the interest rate.
And $D_i=1 (\text{or}~0)$ denotes loan $i$ is successfully funded (or unsuccessful).
For every loan $i$, we construct two variables: the repayment ratio $\lambda_i$ and the return rate $Y_i$.
The repayment ratio $\lambda_i\in[0,1]$ is defined as the ratio of the borrower's successful repayment amount to the total amount the borrower should repay, including the principal and the interest.
$\lambda_i < 1$ means the borrower defaults and $\lambda_i=1$ means the borrower does not default.
The return rate $Y_i\in[0,1+R_i]$ is defined as the ratio of the borrower's successful repayment amount to the principal.
$Y_i=0$ corresponds to the complete loss of the principal and interest; $Y_i=1$ corresponds to the borrower repaying the principal but not the interest; and $Y_i=1+R_i$ corresponds to the borrower repaying the principal and interest in full.
Importantly, the return rate $Y_i$ admits the following decomposition:
\begin{equation} \label{eq:return_rate_decomposition}
Y_i = \lambda_i \times (1+R_i).
\end{equation}

\subsection{Disparate Treatment and Disparate Impact}\label{sec:define_DI}

We start by discussing disparate treatment (DT), a notion that is also legally based but narrowly recognizes direct discrimination.
From the litigation perspective, the disparate treatment doctrine requires the plaintiff to show the \textit{prima facie} disparities are at least in part motivated by the legally protected attributes.
Then, when the defendant articulates some legitimate non-discriminatory reasons~\cite{McDonnel_v_Green} or ``demonstrate that it would have taken the same action in the absence of the impermissible motivating factor''~\cite{Desert_Palace_v_Costa}, the burden shifts back to the plaintiff to demonstrate that the defendant’s stated reasons are insufficient and pretextual~\cite{donahue1996employment}.
In other words, it is ultimately the plaintiff's burden to show the legally protected attributes indeed enter the defendant's decision making.
Statistically, it requires examining whether the protected attributes still have a significant effect on decisions, after controlling all other relevant factors.
The previous literature on P2P lending~\cite{barasinska2014crowdfunding,chen2020gender,pope2011s,ravina2019love,chen2017gender} follows this approach in testing disparate treatment.

Disparate impact has distinct legal elements compared to disparate treatment.
The disparate impact doctrine was formalized in the landmark U.S. Supreme Court case \textit{Griggs v. Duke Power Co.}~\cite{Griggs_v_Duke}.
The Court concluded that the company's facially neutral standardized testing requirement is illegal because (\textit{i}) the tests prevented a disproportionate number of African-American employees from transferring to higher-paying jobs and (\textit{ii}) they are not reasonable measures of job performance.
From the litigation perspective, a plaintiff bringing a disparate impact case bears the initial burden to show both a disparate adverse effect on a protected group and a specific policy that causes such statistical disparity.
The defendant, then, must prove the shown disparity can be ``justified as serving a legitimate business goal'' and no less discriminatory alternatives exists~\cite{EECO_v_Metal,Texas_housing_v_inclusive_communities}.

A statistical measure of disparate impact, therefore, should measure any unwarranted disparity that does not commensurate with individuals' true qualifiedness, irrespective of whether it is direct discrimination.
In P2P lending context, since the investors can determine their own investment amount to every loan, the loan's qualifiedness to funding solely refers to the \textit{return rate}.
Correspondingly, we define disparate impact as the disparate funding of loans that differ in gender but yield identical return rates.

\begin{defn} \label{def:DI}
Disparate Impact (DI) at a return rate level $y$ is defined as follows:
\begin{equation} \label{eq:def_UGD_1}
DI(y) = \mathbb{E} [D\mid G=m, Y=y] - \mathbb{E} [D\mid G=f, Y=y].
\end{equation}
The average level of DI is given by:
\begin{equation} \label{eq:def_UGD_2}
DI = \mathbb{E}_{y}[\Delta(y)],
\end{equation}
where the expectation is taken over the population distribution of the return rate $y$.
\end{defn}

We acknowledge that two important policy components of disparate impact are missing in our statistical measure. 
One is the plaintiff's burden to establish a causal relation between a specific policy and the shown statistical disparity.
The other is the defendant's burden to show no other less discriminatory practice exists.
We recognize the significance of these components and encourage future research to consider them in their analyses.

\section{Two-Stage Predictor Substitution}
We develop a two-stage predictor substitution approach to estimate DI from observational data.
Estimating DI poses a challenge because the return rate is only observable for the funded loans.
To overcome this, in the first stage of 2SPS, we take a predictive approach to impute the return rate for the unsuccessful loans.
Then in the second stage, DI can be directly non-parametrically estimated.

\subsection{First Stage: Survival Model}
We leverage the decomposition of the return rate as the product between the repayment ratio and $(1+\text{the interest rate})$ (Eq.~\ref{eq:return_rate_decomposition}), where the interest rate is fully observed.
We construct a survival model to predict the repayment ratio using the observed loan and borrower characteristics.
Then, the predicted repayment ratio is multiplied with $(1+\text{the interest rate})$, which is fully observed, to produce the predicted return rate.

Survival models focus on modelling \emph{the time to the occurrence of some event}~\cite{cleves2008introduction}, such as patient's time-to-death in biomedical studies~\cite{collett2015modelling,panahiazar2015using}, machine's time-to-failure in engineering, and borrower's time-to-default~\cite{narain1992survival,shumway2001forecasting,stepanova2002survival,dirick2017time} in financial context.
We choose survival model as a predictive model for repayment ratio for several reasons.
First, a borrower's time-to-default provides a sufficient statistics for the repayment ratio because the borrowers make repayments in equated monthly installments and the loans have fixed 12-month installment plan.
Second, survival model explicitly characterizes the default risk over time, leading to more accurate predictions than single period classification models~\cite{shumway2001forecasting}.
Lastly, survival model automatically addresses right-censoring of the borrower's payment record, which occurs in our data due to data cut-off.

Formally, the survival model focuses on $T\in\mathbb{N}=\{0,1,2,\cdots\}$, the random variable that represents the number of months till the occurrence of default.
We call $T$ default time for short.
In our context, $T=0$ means the borrower defaults at the \nth{1} month's installment (and all subsequent months'), and thus corresponds to repayment ratio $\lambda=0$.
$T=1$ means the borrower defaults at the \nth{2} month (and all subsequent months'), and has repayment ratio $\lambda=1/12$.
Similar is true for $T\in\{2,\cdots,11\}$.
$T\geq12$ means the borrower does not default and has repayment ratio $\lambda=1$.

A survival model is completely defined by the hazard function,
\begin{equation} \label{eq:define_hazard_rate}
h(t) = P(T=t\mid T\geq t),
\end{equation}
which defines the instantaneous rate of default given that the borrower has not defaulted before and including the $t$-th month.
Since we consider \emph{monthly} installment and $T\in\mathbb{N}$, $h(t)$ is the probability that the borrower defaults at the $(t+1)$-th month condition on he/she has not defaulted before and including the $t$-th month.
With a slight abuse of language, we call $h(t)$ the hazard rate at the $t$-th month (rather than at the $(t+1)$-th month).

We start with the classical Cox proportional hazard (PH) model~\cite{cox1972regression}, which assumes the following hazard function,
\begin{equation} 
h(t\mid \mathbf{X},\boldsymbol{\beta}) = h_{0}(t) \times \exp \left( \boldsymbol{\beta} \mathbf{X} \right).
\end{equation}
The baseline hazard $h_{0}(t)$ is non-parametric and describes the effect of time for individuals with $\mathbf{X}=\mathbf{0}$, who serves as a reference cell.
The parametric component $\exp(\boldsymbol{\beta} \mathbf{X})$, then, describes the relative increase or decrease of hazard associated with $\mathbf{X}$.
The Cox PH model imposes two strong assumptions---the log-linearity assumption and the proportional hazard assumption---that are unlikely to strictly hold in reality.
We generalize the model to relax these assumptions.

The Cox PH model is a log-linear model, where the continuous covariates act exactly linearly on the log-hazard.
Our first generalization is to allow non-linear relationships by applying natural spline transformation~\cite{therneau2000cox},
\begin{equation} \label{eq:cox_ph_model}
h(t\mid \mathbf{X},\boldsymbol{\beta}) = h_{0}(t)\times \exp(\boldsymbol{\beta} \cdot f_{ns}(\mathbf{X})),
\end{equation}
where $f_{ns}$ denotes the natural spline transformation.
In implementation the natural spline transformation is only applied on the \emph{continuous} covariates.

The proportional hazard (PH) assumption is the distinguishing feature of the Cox PH model.
It, nonetheless, restricts that all loans' hazard rates over time are of a common shape, determined by $h_0(t)$, and the covariates $\mathbf{X}$ affect the hazard rate \emph{time-independently}.
Our second generalization is to allow \emph{time-dependent} effects by adding time interactions,
\begin{equation} \label{eq:generalized_Cox_PH}
h(t\mid \mathbf{X},\boldsymbol{\beta}) = h_{0}(t)  \exp\left( \boldsymbol{\beta}^{(1)} f_{ns}(\mathbf{X}) + \boldsymbol{\beta}^{(2)} f_{ns}(\mathbf{X})  f_{ns}(t) \right),
\end{equation}
where both the covariates and time are natural spline transformed.

We explain how the survival model of Eq.~\ref{eq:generalized_Cox_PH} is fitted in Appendix~\ref{sec:fit_survival_model}, following standard practices in survival analysis~\cite{machin2006survival,cleves2008introduction}.
Using the fitted survival model, we obtain predicted repayment ratio by sampling from the predicted hazard rate over the loan term, with pseudocode shown in Procedure~\ref{procedure:predict_repayment_ratio} in Appendix.
Finally, the predicted repayment ratio is multiplied with $(1+\text{interest rate})$ to produce the predicted return rate.
We acknowledge that more sophisticated survival models, such as mixed cure models~\cite{peng2021cure}, and machine learning models~\cite{wang2019machine} can potentially improve the predictive power.
But we balance between model's simplicity and predictive power.

\subsection{Second Stage: Non-Parametric Estimate}

Using the predicted return rate for the unsuccessful loans as well as a small portion of successful loans ($<3\%$) whose payment record are right-censored due to data cut-off, DI of Eq.~\ref{eq:def_UGD_1} and~\ref{eq:def_UGD_2} can be directly non-parametrically estimated.
To address the challenge of small sample sizes at unique return rate values, we divide the return rate into small intervals and assume a constant loan success rate within each interval.
We obtain confidence intervals by bootstrapping both stages for 500 times.
In every iteration, we resample the data, fit the survival model, and estimate DI using the newly fitted survival model.

\begin{figure*}[!t]
\begin{minipage}{.31\textwidth}
\begin{figure}[H]
\includegraphics[width=\textwidth]{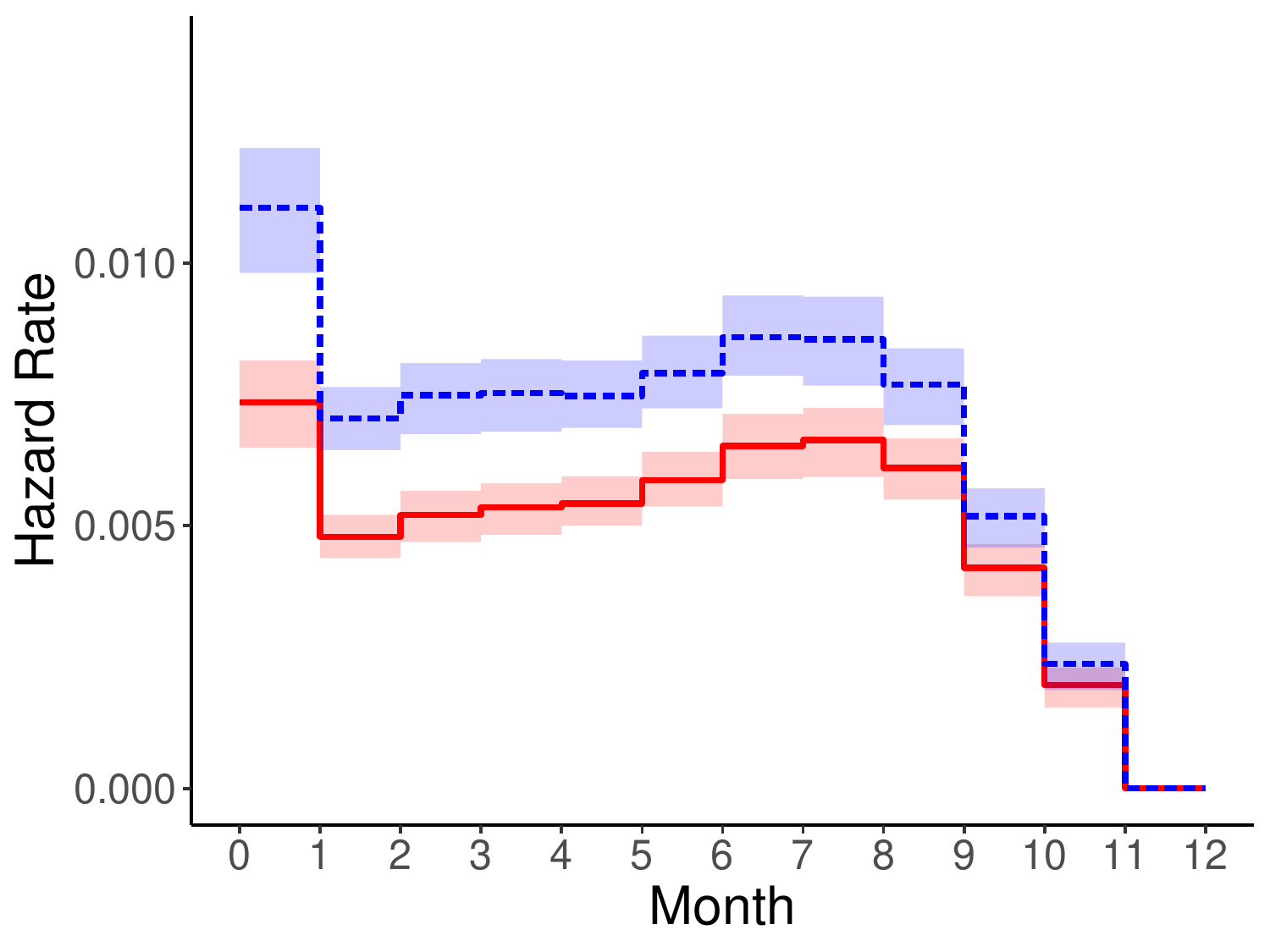}
\caption{Fitted hazard curve with $95\%$ CIs of two loans that have identical covariates $\mathbf{X}$ except borrower gender.
\textcolor{blue}{Blue dashed line} denotes male and \textcolor{red}{red solid line} denotes female.
} \label{fig:fitted_hazard}
\end{figure}
\end{minipage}
\hfill
\begin{minipage}{.31\textwidth}
\begin{figure}[H]
\includegraphics[width=\textwidth]{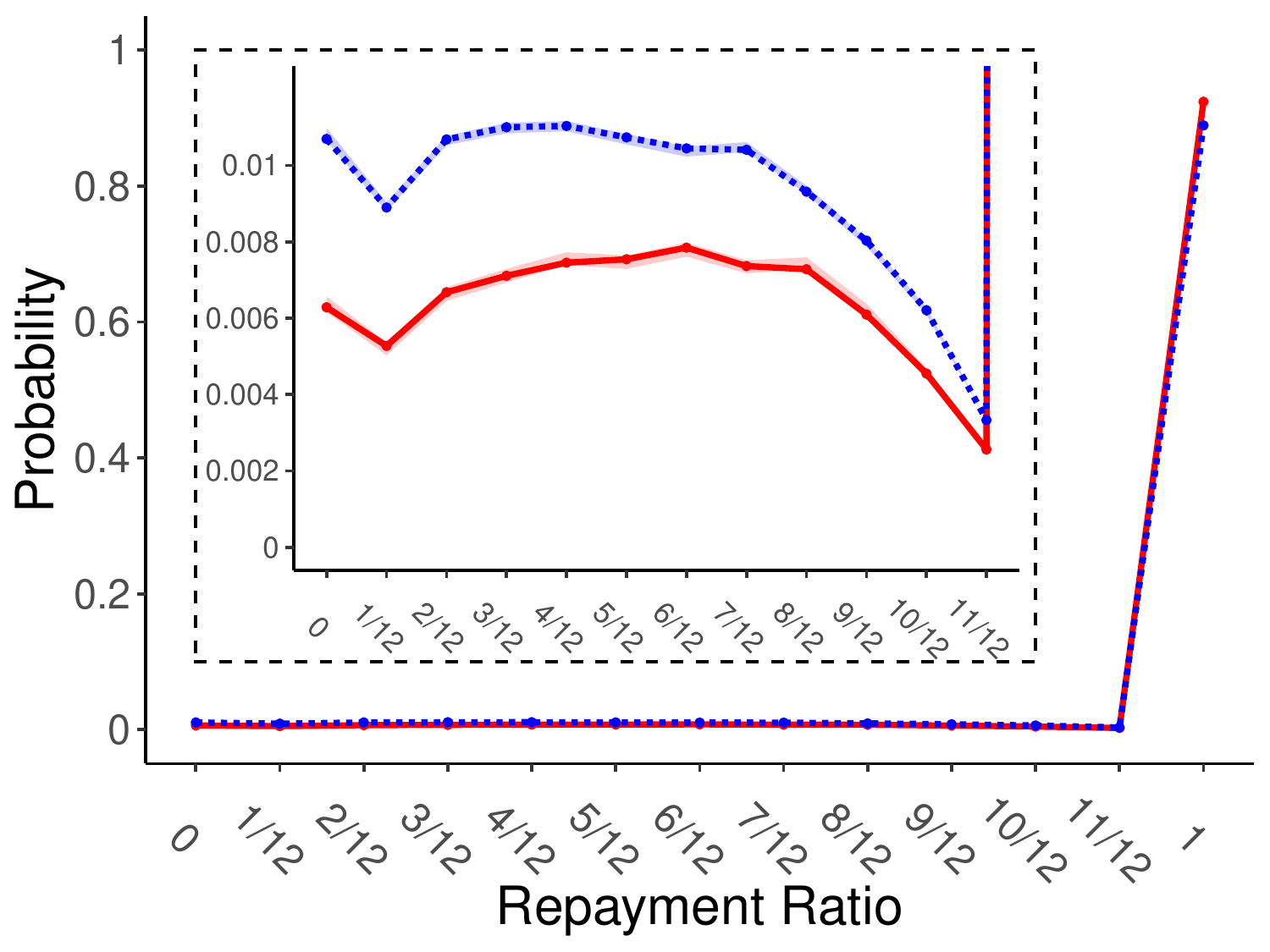}
\caption{Male (\textcolor{blue}{blue dashed line}) and female (\textcolor{red}{red solid line}) borrowers' repayment distribution, using predicted repayment ratio for the unsuccessful loans. Dashed box is the zoom-in view.} \label{fig:fitted_repayment_ratio_distribution}
\end{figure}
\end{minipage}
\hfill
\begin{minipage}{.31\textwidth}
\begin{figure}[H]
\includegraphics[width=\textwidth]{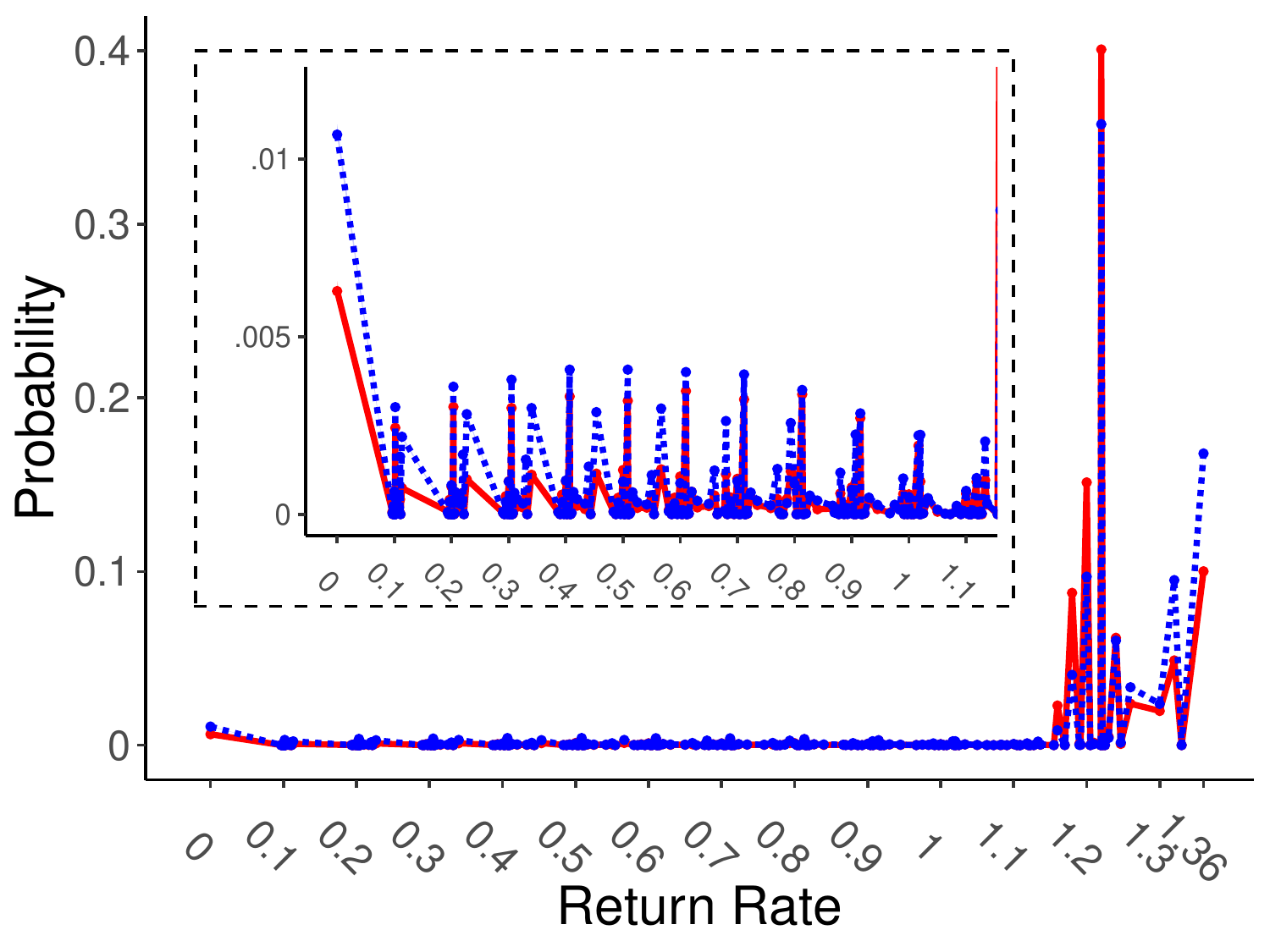}
\caption{Male (\textcolor{blue}{blue dashed line}) and female (\textcolor{red}{red solid line}) borrowers' return rate distribution, using predicted return rate for the unsuccessful loans. Dashed box is the zoom-in view.} \label{fig:fitted_return_rate_distribution}
\end{figure}
\end{minipage}
\end{figure*}

\subsection{Analysis of Bias}
\label{sec:analysis_of_bias}

Using the Bayes rule, DI can be expressed as follows,
\begin{equation}
\begin{aligned}
DI(y) =& \frac{ P_m(D=1) P_m(\lambda(1+R)=y\mid D=1) }
 {\sum_{d\in\{0,1\}} P_m(D=d) P_m(\lambda(1+R)=y\mid D=d) }  \\
&-\frac{ P_f(D=1) P_f(\lambda(1+R)=y\mid D=1) }
 {\sum_{d\in\{0,1\}} P_f(D=d) P_f(\lambda(1+R)=y\mid D=d) } ,
\end{aligned}
\end{equation}
where we use subscript $g$ in $P_g$ to denote the distribution is additionally condition on borrower gender $G=g$.
We use $\hat{DI}(y)$ to denote our 2SPS estimate of $DI(y)$, which uses predicted repayment ratio $\hat{\lambda}$ for the unsuccessful loans\footnote{For ease of analysis, we ignore the small fraction ( $<3\%$) of successful loans whose payments are partially unobserved due to data cutoff.
},
{\small
\begin{equation}
\begin{aligned}
&\hat{DI}(y) = \\
& 
\frac{ P_m(D=1) P_m(\lambda(1+R)=y\mid D=1) }
 {
 \left(\makecell{P_m(D=1) P_m(\lambda(1+R)=y\mid D=1) \\+ P_m(D=0)\sum_{r}P_m(R=r\mid D=0) \textcolor{red}{P_m(\hat{\lambda}=y/r\mid D=0, R=r) }}\right)} \\ 
 &-\frac{ P_f(D=1) P_f(\lambda(1+R)=y\mid D=1) }
 {
 \left(\makecell{P_f(D=1) P_f(\lambda(1+R)=y\mid D=1) \\+ P_f(D=0)\sum_{r}P_f(R=r\mid D=0) \textcolor{red}{P_f(\hat{\lambda}=y/r\mid D=0, R=r) }}\right)}  .
\end{aligned}
\end{equation}
}

We define bias from the first stage of 2SPS as $b_{g,r}(l)$,
\begin{equation} \label{eq:def_first_stage_bias}
b_{g,r}(l) =P_g(\hat{\lambda}=l \mid D=0, R=r) - P_g(\lambda=l \mid D=0, R=r).
\end{equation}
The first-stage bias $b_{g,r}(l)$ is measured at a specific repayment ratio level $l$ condition on borrower gender $g$, interest rate $r$, and loans being unsuccessful $D=0$.
We now can express the asymptotic bias for our 2SPS estimate of DI in terms of the first-stage bias:
{\small
\begin{equation} \label{eq:2SPS_bias}
\begin{aligned}
b_{2SPS}(y) =& \hat{DI}(y) - DI(y) \\
=& \frac{\textcolor{blue}{P_f(D=0)} \times \overbrace{\textcolor{red}{\sum_{r} P_f(R=r\mid D=0)b_{f,r}(y/r)}}^{\text{average female first-stage bias $b_{f}$}}}{
\textcolor{blue}{
\left(1+\frac{P_f(\hat{\lambda}(1+R)=y,D=0)}{P_f(\lambda(1+R)=y,D=1)}\right) 
\left(1+\frac{P_f(\lambda(1+R)=y,D=0)}{P_f(\lambda(1+R)=y,D=1)}\right)
}} \\
&- \frac{\textcolor{blue}{P_m(D=0)} \times \overbrace{\textcolor{red}{\sum_{r} P_m(R=r\mid D=0)b_{m,r}(y/r)}}^{\text{average male first-stage bias $b_{m}$}}}{\textcolor{blue}{\left(1+\frac{P_m(\hat{\lambda}(1+R)=y,D=0)}{P_m(\lambda(1+R)=y,D=1)}\right) \left(1+\frac{P_m(\lambda(1+R)=y,D=0)}{P_m(\lambda(1+R)=y,D=1)}\right)}} .
\end{aligned}
\end{equation}
}

Therefore, our 2SPS estimate is unbiased as long as $b_{2SPS}(y)=0$.
A more strict but sufficient condition is both the average first-stage biases $b_f$, $b_m$ are zero.
Two comments are worth highlighting.
First, as defined in Eq.~\ref{eq:def_first_stage_bias}, the first-stage bias $b_{g,r}(l)$ is measured \emph{without conditioning on $\mathbf{X}$}.
We only need the predicted repayment ratio $\hat{\lambda}$ to induce a distribution unbiased to $P_g(\lambda \mid D=0, R=r)$, which is a much weaker condition than unbiasedness of the predicted repayment ratio itself, $\mathbb{E}[\hat{\lambda}\mid \mathbf{X}] = \mathbb{E}[\lambda\mid \mathbf{X}] $.
Second, in $b_{2SPS}(y)$ the first-stage bias $b_{g,r}(y/r)$ is further marginalized over the interest rate $r$ (\textcolor{red}{red terms}).
\textit{I.e.}, only the \emph{average} first-stage biases affect the 2SPS estimate.

The 2SPS estimate is also robust to the average first-stage biases $b_f$, $b_m$ when they are non-zero.
The reasons are as follows.
First, $b_f$ and $b_m$ are both down-scaled in $b_{2SPS}(y)$: the coefficient in front of them (\textcolor{blue}{blue terms}) are always smaller than 1.
Second, When $b_f$ and $b_m$ have the same sign, they partially cancel each other in $b_{2SPS}(y)$.
Intuitively, the 2SPS estimate is partially protected from systematic over- or underestimation of repayment ratio.

\section{Empirical Results}

\subsection{Survival Model}
We plot the fitted hazard curve in Fig.~\ref{fig:fitted_hazard} and report the survival model's fitted coefficients in Tab.~\ref{tab:survival_categorical_coefficients} and Fig.~\ref{fig:survival_term_plot} in Appendix.
For both male and female borrowers, the hazard rate is the highest at the \nth{1} month, drops abruptly at the \nth{2} month, and then slightly increases till around the \nth{8} month, and finally decreases to around $0$ at the \nth{12} month.
Our interpretation is: the uncollateralized nature of online P2P lending invites a number of ill-intentioned borrowers who intend to default all installments.
Fewer borrowers default at the last several installments because it is uneconomical: they could either default at an earlier time to earn a higher economic gain, or do not default at all to maintain their credit.
Notably, male borrowers have an increased hazard rate, 1.503 (log-S.E.=0.021) times that of the female borrowers.
Female has a lower propensity to default at all months.

We report and discuss diagnostics of the fitted survival model in Appendix~\ref{sec:survival_model_diagnostics}.
We assess the proportional hazard (PH) assumption by plotting the scaled Schoenfeld residual~\cite{schoenfeld1982partial,grambsch1994proportional}.
We assess the model's goodness-of-fit by plotting the Cox-Snell residual~\cite{cox1968general}.
And we report the fitted survival model's predictability in terms of the concordance index~\cite{harrell1996multivariable}.
Results show the survival model is well-specified, has goodness-of-fit, and is predictive of borrowers' default.
We note these are standard and widely accepted techniques for assessing the adequacy of survival models and we refer the readers to Section 4 of~\citet{collett2015modelling} for a detailed review.

\begin{figure*}[t!] 
\includegraphics[width=.48\textwidth]{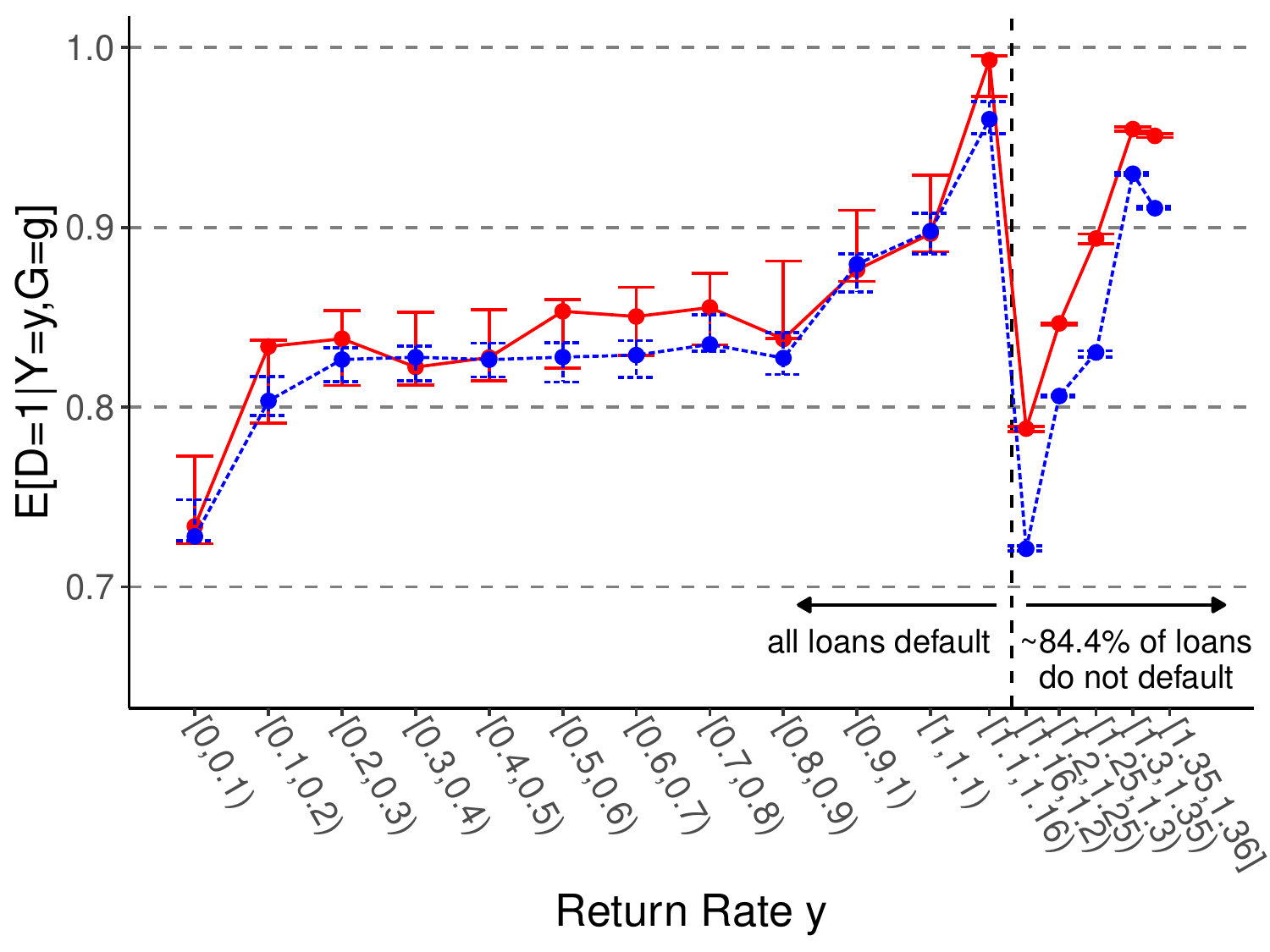}
\hfill
\includegraphics[width=.48\textwidth]{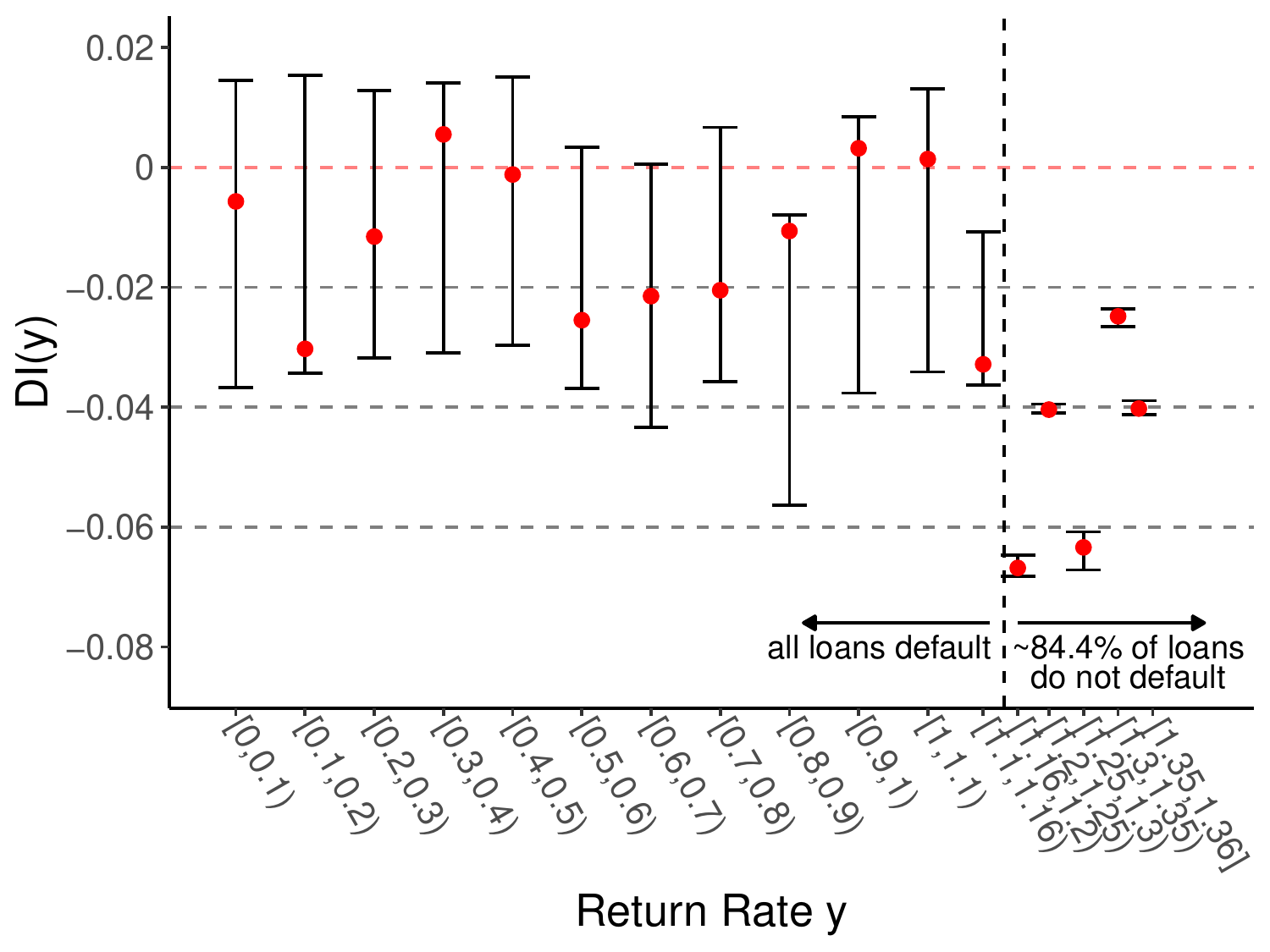}
\caption{Left and right figures plot the estimated conditional loan success rate and disparate impact (DI), respectively.
In the left figure, \textcolor{red}{red solid line} denotes female and \textcolor{blue}{blue dashed line} denotes male.
Since the interest rate in our data ranges within $[0.16,0.36]$, all loans with return rate $<1.16$ are defaulted loans.
Around $84.4\%$ of loans with return rate $\geq1.16$ do not default.
} \label{fig:gender_discrimination_by_return_rate}
\end{figure*}

\begin{table*} \centering 
\begin{tabular}{c | r c c c } 
\hline
 \textit{2SPS w/ non-parametric 2nd stage} & & \multicolumn{3}{c}{\textit{2SPS w/ OLS regression 2nd stage}} \\ 
\cline{1-1} \cline{3-5} 
Disparate Impact (DI) & & (1) DI & (2) DI, Control $\mathbf{X}$. & (3) Disparate Treatment (DT) \\ 
\hline 
 -0.0397 & Male & -0.0388 & -0.0244 & -0.0215  \\ 
  (-0.0398$, $-0.0395) & & (-0.0389$, $-0.0385) & (-0.0245$, $-0.0242) & (-0.0234$, $-0.0196)  \\ 
  & Return Rate $Y$ & \checkmark & \checkmark \\
  & Loan \& Borrower Charc. $\mathbf{X}$ & & \checkmark & \checkmark \\
\hline 
\end{tabular} 
\caption*{Various 2SPS estimates.
Some covariates in $\mathbf{X}$ are eliminated in the backward feature selection procedure of OLS regression using Akaike Information Criterion (AIC)~\cite{akaike1974new}.}
\label{tab:UGD_OLS} 
\end{table*}

\subsection{Repayment Ratio and Return Rate}

We plot the gender-specific repayment ratio distribution and return rate distribution in Fig.~\ref{fig:fitted_repayment_ratio_distribution} and~\ref{fig:fitted_return_rate_distribution}, using the predicted values for the unsuccessful loans.
We find female borrowers' repayment ratio distribution \emph{dominates} the male borrowers'.
A higher fraction of female borrowers---$92.4\%$ compared to $88.9\%$ for male borrowers---do not default and have repayment ratio $\lambda=1$.
And fixing any repayment ratio level $\lambda\in\{0,1/12,\cdots,11/12\}$ where default does occur, there is always a lower fraction of female compared to male.
Nonetheless, the female borrower' return rate distribution does not dominate the male borrower's.
Female's return rates are more concentrated in an intermediate range, between $1.16\sim1.24$.
Male's return rate has larger variation: their loans are more likely to yield both very high ($>1.24$) and very low ($<1.16$) return rates.
The reason is, as reported in the descriptive statistics (Tab.~\ref{tab:descriptive_table}), males on average borrow at a higher interest rate, which partially compensate their lower repayment ratio.

\subsection{Disparate Impact}

Fig.~\ref{fig:gender_discrimination_by_return_rate} reports the estimated loan success rate and disparate impact.
Let's first focus on the loan success rate.
The trend that loans with higher return rates are more likely to be funded holds \emph{individually within the loans whose return rate $<1.16$ and within the loans whose return rate $\geq1.16$}.
The loan success rate drops abruptly from the return rate interval $[1.1,1.16)$ to $[1.16,1.2]$.
Our explanation is while these two set of loans yield similar return rates, they differ greatly in their interest rates.
Since the interest rate in our data ranges between $[0.16,0.36]$, the loans with return rate between $[1.16,1.2]$ are mostly the loans that do not default but offer some of the lowest interest rates on the platform\footnote{To be precise, these are the loans with the lowest interest rates the investors actually consider lending to. During the data pre-processing stage, we exclude the loans with interest rates below $16\%$ because they are never funded.}.
In contrast, the loans with return rate between $[1.1,1.16]$ are the loans that offer higher interest rates but later defaulted.
Therefore, the investors appear to be risk-seeking in funding loans with return rate $[1.1,1.16)$ at a much higher rate than loans with return rate $[1.16,1.2)$.

DI is estimated as the difference between male and female's loan success rates, conditional on identical return rates.
We observe marginal DI favoring female for loans with return rate $<1.16$ but very significant DI favoring female---with magnitude varing between 2.48\% and 6.68\%---for loans with return rate $\geq 1.16$.
The CIs are especially sharp for the latter because around $90\%$ of the loans in our data have return rate $\geq1.16$.
Averaged over the population distribution of return rate, we find female borrowers are \textbf{3.97\% (95\% CI: -3.98\%$\sim$-3.95\%)} more likely to have their loans successfully funded, given identical return rates.
We note that as reported in the descriptive statistics (Tab.~\ref{tab:descriptive_table}), female borrowers \textit{prima facie} have 2.2\% higher loan success rate, 85.9\% compared to 83.7\% for male borrowers.
By additionally controlling the return rate, DI reveals the female favoritism on the platform is around 1.8 times the \textit{prima facie} gender difference.

\subsection{Analysis}
\label{sec:analysis_DI}

\noindent\textbf{Indirect and Proxy Discrimination}\quad
We use a decomposition technique to investigate how much of DI is indirect and proxy discrimination.
This procedure proceeds as follows.
We first replace the second stage of 2SPS with an OLS regression, regressing loan success on borrower gender and return rate.
The return rate is spline transformed to allow non-linear effects.
The gender coefficient gives a different 2SPS estimate of DI where the second stage is parametric.
Then, by additionally controlling the observed loan and borrower characteristics in OLS regression, we measure the \emph{reduction of the gender effect}, which identifies the DI that can be explained by them.
This estimate gives a \emph{lower bound} for indirect and proxy discrimination, since there might be unobserved mediators and proxy variables.
In a concurrent work, \citet{bohren2022systemic} develop a similar decomposition-based approach to distinguish direct and systemtic discrimination.

The results are reported in Tab.~\ref{tab:UGD_OLS}.
First, the OLS estimate finds female is $3.88\%$ more likely to be funded given identical return rates.
It is very close to and corroborates our original 2SPS estimate of $3.97\%$.
Second, additionally controlling for observed loan and borrower characteristics significantly reduces the gender effect to $2.44\%$.
It means at least $37.1\%$ of the DI favoring female is indirect or proxy discrimination.
Lastly, we also report in Tab.~\ref{tab:UGD_OLS} an estimate of DT, which controls for loan and borrower characteristics but \emph{not the return rate}.
DT indeed underestimates the female favoritism by $44.6\%$, equivalent to saying female is $2.15\%$ more likely to be funded than male, \textit{ceteris paribus}.

\noindent\textbf{Data Disaggregation}
To investigate how DI varies in different cohorts of loans or borrowers, we report in Tab.~\ref{tab:UGD_disaggregated} in Appendix the reestimated DI in data subsets disaggregated by borrowing time, marriage, repeated borrower, age, borrowing amount, interest rate, employment, and education.
We find consistent DI favoring female in almost all subsets, with mild variation in its magnitude.
The DI favoring female is the largest for student borrowers at 6.45\%, and is larger for loans with higher interest rates.
Notably, we find the magnitude of DI favoring female decreases with higher education, and even becomes \emph{against female} for borrowers with Master or Doctorate degrees.
But these borrowers only constitute around $0.2\%$ of our data.
This is the only case where we find DI becomes against female.

\noindent\textbf{Sensitivity to Overestimation of the Repayment Ratio}
In the 2SPS approach, we predict the repayment ratio for the unsuccessful loans using data from the successful loans.
The concern is we might systematically overestimate the unsuccessful loan's repayment ratio because the investors use information that is unrecorded but is indeed predictive of default in their lending decisions.
One such concrete example is the loan's description, which the investors do observe, is not available in our data, and is reasonably believed to be predictive of default~\cite{jiang2018loan,xia2020predicting}.
This issue is generally unavoidable as it is unrealistic to account for all factors that are predictive of default.
Although Section~\ref{sec:analysis_of_bias} theoretically shows the 2SPS estimate is partially protected from systematic overestimation of the repayment ratio, we are still interested in empirically testing its sensitivity.

Thus, we simulate the scenario where there is an omitted fixed effect associated with the unsuccessful loans.
A fixed effect in the survival model is a constant multiplicative factor to the hazard rate (defined in Eq.~\ref{eq:define_hazard_rate}) across all months.
We assume the omitted fixed effect is in the direction that the actual hazard rates are higher than our current predictions, and vary its strength from $\{1.5,2,2.5,3\}$.
Concretely, the unsuccessful loans' predicted hazard rates across all months are multiplied with 1.5, 2, 2.5, or 3.

\begin{figure}[t!] 
\vspace{-3ex}
\centering
\includegraphics[width=.8\columnwidth]{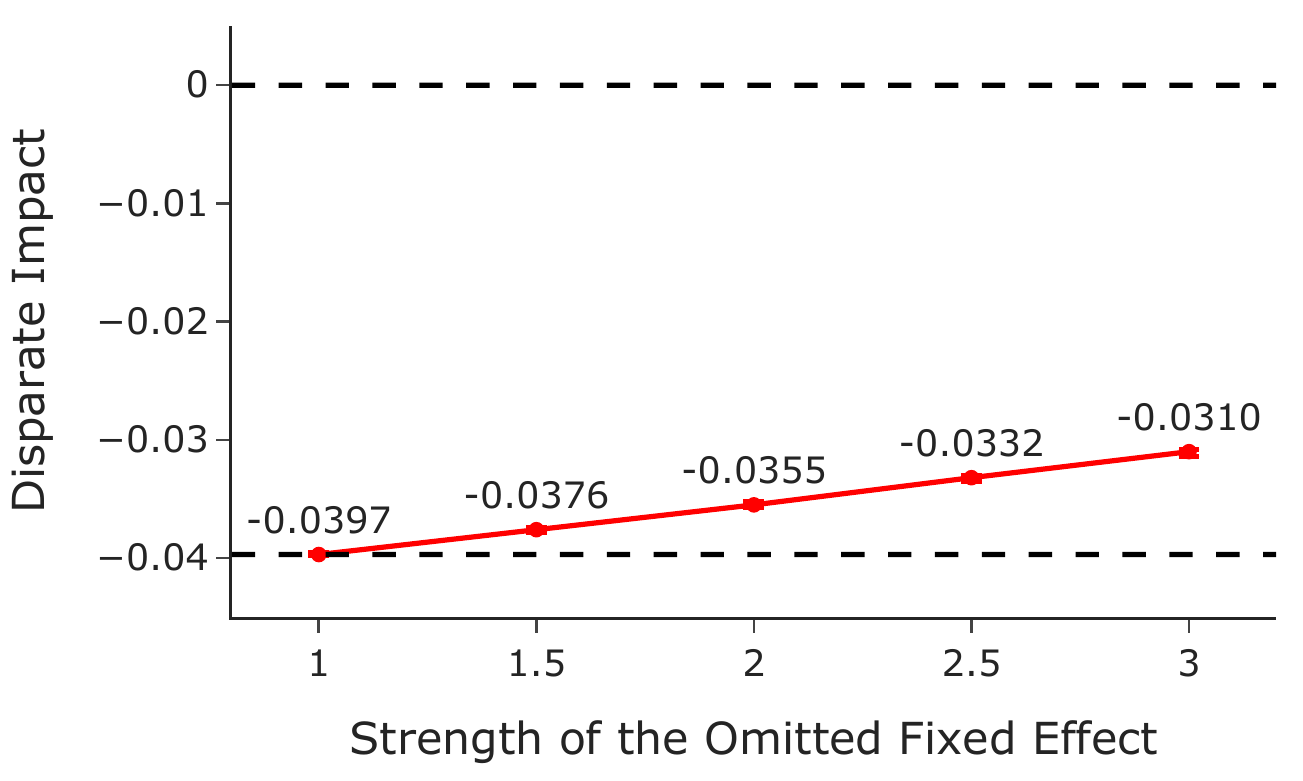}
\caption{Disparate Impact (DI) Under Simulated Overestimation of Repayment Ratio.} \label{fig:gender_discrimination_under_omitted_effect}
\vspace{-2.5ex}
\end{figure}

The results are reported in Fig.~\ref{fig:gender_discrimination_under_omitted_effect}.
We observe a mild linear decrease in the magnitude of DI favoring female.
The DI favoring female is still more than $3\%$ when the unsuccessful loans' default risks are underestimated by three-fold.
Applying a linear extrapolation, if the actual DI is zero or is \emph{against} female, the unsuccessful loans' actual hazard rates will have to be \emph{at least $10$ times our current predictions}.
Using Fig.~\ref{fig:fitted_hazard} as a reference, it means their hazard rates are at least around 5\% at \emph{every month except the very last months}.
We argue this scenario is extremely unlikely.
Thus, our finding of DI favoring female is robust to potential overestimation of repayment ratio.

\begin{table*}[!t]
\centering
\begin{tabular}{l c c c c c c}
\toprule
 & \multicolumn{4}{c}{Statistical Disc.} & & Taste-based Disc. \\
 \cline{2-5} \cline{7-7}
  &  \makecell{Repayment Ratio\\Mean $\mu_g$} & \makecell{Repayment Ratio\\Variance $\sigma_{g,0}$} & \makecell{Repayment Ratio\\Noise Variance $\sigma_{g,1}$} & \makecell{Signal\\Reliability $\gamma_g$} && \makecell{Funding\\Threshold $\pi_g$} \\
 \midrule
 \multirow{2}{*}{Female} & $0.957$ & $0.167$ & $0.482$ & $0.107$ && $1.099$ \\
 & $(0.956\sim0.957)$ & $(0.166\sim 0.168)$ & $(0.477\sim0.487)$ & $(0.105\sim0.109)$ && $(1.098\sim1.100)$ \\
 \multirow{2}{*}{Male} & $0.934$ & $0.205$ & $0.574$ & $0.113$ && $1.079$ \\
 & $(0.934\sim0.935)$ & $(0.205\sim0.206)$ & $(0.571\sim0.578)$ & $(0.112\sim0.114)$ && $(1.078\sim1.080)$ \\
\bottomrule
\end{tabular}
\caption{Estimated Parameters from The Decision Model.}  \label{tab:estimated_true_decision_parameters}
\end{table*}

\section{Taste-based and Statistical Discrimination}

\subsection{Decision Model}
\label{sec:decision_model}

Next, we investigate a decision model which decomposes DI into two components due to taste-based~\cite{becker1971economics} and statistical discrimination~\cite{phelps1972statistical,arrow1973theory,aigner1977statistical}.
Concretely, we assume every loan $i$'s repayment ratio is sampled from a gender-specific Gaussian distribution, $\lambda_i \sim N(\mu_g, \sigma_{g,0}^2)$, with repayment ratio mean $\mu_g$ and repayment ratio variance $\sigma_{g,0}$.
We stylize the process of investors evaluating various information of the loan as perceiving a noisy repayment ratio signal $\hat{\lambda}_i$, $\hat{\lambda}_i = \lambda_i + \sigma_{g,1}\epsilon_i, \epsilon_i\sim N(0,1)$, with gender-specific noise variance $\sigma_{g,1}$.
We assume the investors are able to accurately predict the expected repayment ratio $\tilde{\lambda}_i$,
\begin{align*}
\tilde{\lambda}_i\mid (G_i=g) =& \mathbb{E}[\lambda_i\mid \hat{\lambda}_i] = (1-\gamma_g) \mu_g + \gamma_g (\lambda_i + \sigma_{g,1} \epsilon_i), \\
\gamma_g =& \frac{(\sigma_{g,1})^{-2}}{(\sigma_{g,0})^{-2}+(\sigma_{g,1})^{-2}},
\end{align*}
where $\gamma_g$ is known as the signal reliability. 
$\gamma_g$ exactly captures the repayment ratio variance $\sigma_{g,0}$ and the repayment ratio noise variance $\sigma_{g,1}$'s effect in computing the expected repayment ratio $\tilde{\lambda}_i$.
The above process formalizes statistical discrimination, where the investors leverage the easily observable borrower gender to accurate predict the expected repayment ratio $\tilde{\lambda}_i$. But doing so might have a disparate impact on male and female borrowers.

Then, since the interest rate $R_i$ is fully observed, we assume the loan's funding success is determined by whether the expected return rate, computed as the product between the expected repayment ratio $\tilde{\lambda}_i$ and $(1+R_i)$, exceeds a gender-specific threshold $\pi_g$,
$$
D_i\mid (G_i=g) = \mathds{1}[\tilde{\lambda}_i(1+R_i) \geq \pi_g].
$$
The funding threshold $\pi_g$ formalizes taste-based discrimination.
The investors are said to have an intrinsic gender taste that favors (or disfavors) female if $\pi_f<\pi_m$ (or $\pi_f>\pi_m$).
We will use gender taste, gender animus, and taste-based discrimination interchangeably.

Two comments are in order.
First, this model gives the investors the benefit of the doubt by assuming they accurately predict the expected repayment ratio.
While it is likely the investors are inaccurate, inaccurate statistical discrimination can be observationally equivalent to gender taste~\cite{arnold2018racial,hull2021marginal}.
We therefore follow the common approach in regarding inaccurate statistical discrimination as also arising from gender taste~\cite{arnold2020measuring}.
Second, this model gives an ``as-if'' characterization of the investors' collective funding behavior, but does not imply their actual intentions, which is a much more difficult task.
This issue is inherent to most works inferring discrimination drivers from observational data~\cite{hull2021marginal,simoiu2017problem}.
The findings from this model is still valuable, for example in ruling out statistical discrimination as the only driver of disparate impact.

We estimate parameters from this model using a similar 2SPS approach.
The first stage is identical: a survival model is fitted to predict the repayment ratio for the unsuccessful loans.
This directly allows us to estimate the repayment ratio mean $\mu_g$ and variance $\sigma_{g,0}$.
In the second stage, using the predicted repayment ratio and the estimated $\mu_g$ and $\sigma_{g,0}$, we use Bayesian inference to infer the posteriors of the repayment ratio noise variance $\sigma_{g,1}$ and the funding threshold $\pi_g$.
Finally, the signal reliability $\gamma_g$ can be derived using the estimated $\sigma_{g,0}$ and $\sigma_{g,1}$.
The Bayesian inference step is elaborated in Appendix~\ref{sec:bayesian_inference}.
It has good convergence as indicated by the Gelman-Rubin diagnostic $\hat{R}$~\cite{gelman1992inference} at most 1.0023 for all parameters, as well as visual inspection of the traces shown in Fig.~\ref{fig:Bayesian_traces}.
We bootstrap both stages for 500 times to obtain confidence intervals.
We note the second stage closely resembles a threshold test for discrimination~\cite{simoiu2017problem,pierson2018fast}.

\subsection{Results}

Tab.~\ref{tab:estimated_true_decision_parameters} reports the estimated parameters.
Most importantly, we find (\textit{i}) the observed DI favoring female can be completely explained as rational statistical discrimination and (\textit{ii}) female borrowers still need to have 2\% higher \emph{expected} return rate in order to be funded, \textit{i.e.}, gender animus against female still exists.
To put another way, if the investors only engage in statistical discrimination and do not have taste-based discrimination, female borrowers would be \emph{more favorably funded}.
The statistical discrimination is mainly driven by two factors: (\textit{i}) female borrowers have a higher repayment ratio mean $\mu_g$, $0.957$ compared to $0.934$ for male borrowers; and (\textit{ii}) because the investors' signal reliability $\gamma_g$ is very low, the repayment ratio mean $\mu_g$ plays a significant role in computing the expected repayment ratio $\tilde{\lambda}_i$.

\section{Discussions}

\noindent\textbf{P2P Lending as a Market-based Affirmative Action} \quad
Given the reality that gender gap in financial access still persists~\cite{demirgucc2022global} and female is still disadvantaged in many other credit markets~\cite{wellalage2021bank,chaudhuri2020gender,brock2021discriminatory}, our empirical results suggest online P2P lending provides an alternative credit market where the affirmative action to support female can arise naturally from the rational crowd.
We take a liberal interpretation of affirmative action that focuses on the effect of narrowing existing inequality but does not imply intentions from the investors and/or the platform, similar to the discussion of market affirmative action in~\citet{cooter1994market}.
Below, we discuss two reasons why P2P lending can be a desirable policy instrument to narrow the gender gap in financial access.

First, the affirmative action in P2P lending is driven by market forces, which is arguably more efficient than heavy-handed bureaucratic rules such as quota systems~\cite{beauregard2021antiwomen,clayton2015women}.
The P2P lending market functions like a perfectly competitive market because (\textit{i}) the size of P2P loans is typically small and (\textit{ii}) there is a large number of investors.
Based on standard demand-supply analysis~\cite{cooter1994market}, the market competition will impose the cost of gender animus---in the form of lower expected return---upon the investors who demand it.
Therefore, competition reduces gender animus.
Competition, however, will not eliminate statistical discrimination because it is efficient and reflects rational behavior~\cite{phelps1972statistical}.
Consequently, because females are indeed less likely to default on their P2P loans, the market forces will drive the investors to engage in statistical discrimination that favors female.

Second, affirmative action resulted from statistical discrimination leads to an important narrative shift.
The female borrowers' favorable DI is incentivized by their own higher expected return rates rather than being granted due to their disadvantaged socio-economical status.
Particularly, it generates a viable defence to the attack on affirmative action that less qualified female borrowers are selected over more qualified male borrowers~\cite{buchanan1989johnson}.
In fact, more qualified female borrowers---in terms of \emph{expected return rates}---are selected by the investors.

On a cautionary note, statistical discrimination by itself is likely to be illegal in many jurisdictions. In order to qualify as affirmative action, a practice must be narrowly tailored, not violate the rights of non-protected groups, and meet other legal requirements. 
Since we lack expertise in law, we refrain from delving into the discussion of legality in this paper.

\noindent\textbf{Gender Animus Can Still Exist Under Favorable Disparate Impact} \quad
While our study on disparate impact reveals a larger female favoritism on the collaborated P2P lending platform than disparate treatment, We caution against making overly simplistic or absolute interpretations of the findings. 
We provide counter evidence that the overall female favoritism can be completely explained as rational statistical discrimination and gender animus \emph{against female} can still exist.
Gender animus does not become right or justified when its effect is obscured by statistical discrimination.
The underpinning is different discrimination drivers, such as gender taste and statistical discrimination, can co-exist and contribute to the overall discrimination.
We call for future research that not only tests the existence of discrimination but also identifies the underlying drivers of discrimination, as highlighted by~\citet{hull2021marginal,bohren2019inaccurate}.

\section{Limitations}
There are several limitations in this work, including (1) treating gender as the binary biological sex, (2) the risk of overestimating the unsuccessful loan's repayment ratio due to omitted factors, and (3) the nature of an ``as-if'' characterization in attempting to identify discrimination drivers.
This work is also limited in its inability to discuss individual investors' gender discrimination due to data constraint.
Similarly, we are unable to explicitly investigate the ML-driven credit grade' effect on gender discrimination due to the unavailability of the credit grade information.
This also presents an opportunity for future research.

\section{Conclusion}
This work presents a case study of gender discrimination and its underlying drivers on a prominent Chinese online P2P lending platform.
We measure the disparate impact (DI) favoring female is around 3.97\%, which reveals a larger female favoritism than commonly studied discrimination notion of disparate treatment (DT).
But we also identify this female favoritism can be explained as rational statistical discrimination.
Gender animus against female can still exit and be obscured by other discrimination drivers. 
We conclude by discussing the positive role P2P lending can play to reduce the existing gender gap in financial access and the importance of, besides measuring the overall discrimination, identifying what drives discrimination and decomposing their effects.

\bibliographystyle{ACM-Reference-Format}
\bibliography{reference}

\clearpage
\appendix
\onecolumn

\section{Appendix}
\setcounter{table}{0}
\renewcommand{\thetable}{\Alph{section}\arabic{table}}
\setcounter{figure}{0}
\renewcommand{\thefigure}{\Alph{section}\arabic{figure}}

\subsection{Descriptive Statistics of Our Data} \label{appdix:descriptive_statistics}

{\small
\begin{table}[h]
\centering
 \begin{tabular}{l c c c} 
 \toprule
  & \makecell[c]{All\\Borrower} & \makecell[c]{Male\\Borrower} & \makecell[c]{Female\\Borrower} \\
 \midrule
 \# Loans & 676,473 & 519,099 & 157,374 \\
 \# Successful Loans & 569,879 & 434,739 & 135,140 \\
 Success Rate & 0.842 & 0.837 & 0.859 \vspace{2ex}\\
 \textit{Panel A: Borrower Charc.} \\
 \quad ID Province & - & - & - \\
 \quad Male & 0.767 & - & - \\
 \quad Married & 0.490 & 0.485 & 0.506 \\
 \quad Age & 28.29  & 28.48  & 27.66  \\
 \quad Repeated Borrower & 0.479 & 0.484 & 0.460 \\
 \quad Employment: \\
 \quad \quad (0) Unknown & 0.271 & 0.251 & 0.338 \\
 \quad \quad (1) Student & 0.038 & 0.037 & 0.041 \\
 \quad \quad (2) Worker & 0.490 & 0.506 & 0.438\\
 \quad \quad (3) Self-employed & 0.189 & 0.195 & 0.167 \\
 \quad \quad (4) Online Shop Owner & 0.012 & 0.011 & 0.017 \\
 \quad Education: \\
 \quad \quad (0) Unknown & 0.744 & 0.753 & 0.717 \\
 \quad \quad (1) Junior College & 0.013 & 0.011 & 0.014 \\
 \quad \quad (2) College & 0.159 & 0.152 & 0.181 \\
 \quad \quad (3) Bachelor & 0.082 & 0.081 & 0.087 \\
 \quad \quad (4) Master or Doctorate & 0.002 & 0.001 & 0.001 \\
 \quad \# Past Failed Borrowings & 0.268  & 0.264  & 0.282  \\
 \quad \# Past Aborted Borrowings & 0.373  & 0.387  & 0.327  \\
 \quad \# Past Ontime Payments & 3.740  & 3.818  & 3.484  \\
 \quad \# Past Late Payments & 0.849  & 0.874  & 0.766  \vspace{2ex}\\
 \textit{Panel B: Loan Charc.} \\
 \quad Amount (thousand RMB) & 3.403  & 3.354  & 3.566  \\
 \quad Interest Rate & 0.254 & 0.259  & 0.237  \\
 \quad APP Channel & 0.207 & 0.187 & 0.272 \\
 \quad Express Loan & 0.020 & 0.015 & 0.035 \\
 \bottomrule
 \end{tabular} 
 \caption{This table reports the mean statistics of our data. APP Channel denotes whether the loan application is made through the platform's mobile application. Express Loan is a special category of loans.
 } \label{tab:descriptive_table}
\end{table}
}

\subsection{Fitting Survival Model}
\label{sec:fit_survival_model}
We briefly explain how the survival model is fitted to the data, following standard practices in survival analysis~\cite{machin2006survival,cleves2008introduction}.
We note that for some successful loans, the default time $T$ is unobserved because the borrower does not default or payment records are right-censored.
The latter scenario is indeed present in our data due to data cutoff.
To address this issue, we implicitly express $T$ using two other variables, the observation time $\tau\in\{0,1,\cdots,12\}$ and the censoring indicator $C\in\{0,1\}$~\cite{collett2015modelling,kleinbaum2012survival}.
Loans that do not default have observation time $\tau=12$ and censoring indicator $C=1$ because we have observed the loan for 12 months without default occuring and our observation is censored right after the 12-th month, indicating default time $T\geq12$.
Defaulted loans have observation time $\tau=T$ and censoring indicator $C=0$.
The loans whose payment record is right-censored have their respective observation time $\tau$ and censoring indicator $C=1$.
In this expression, all successful loans have an observation time $\tau$ and a censoring indicator $C$.
Then, a partial likelihood can be derived using the counting process formulation~\cite{aalen1978nonparametric,andersen1982cox,auton2000applied} and the Elfron approach~\cite{efron1977efficiency}.
We finally obtain the empirical maximum partial likelihood estimate of the survival model parameters.

\subsection{Predict Repayment Ratio Using Fitted Survival Model}

\begin{algorithm}[H]
\SetAlgorithmName{Procedure}{Procedure}{Procedure}
\SetKwInOut{Input}{input}
\caption{Predict Repayment Ratio $\lambda$} \label{procedure:predict_repayment_ratio}
\Input{ 
$\hat{\boldsymbol{\beta}}$, fitted survival model; 
$\mathbf{X}$, covariates; 
$t_0\in\{0,1,\cdots,11\}$, installments are successfully repayed before and including the $t_0$-th month, but are unobserved after and including the ($t_0+1$)-th month.}
$t \gets t_0$, $Alive \gets \text{True}$\;
\While{$t\leq 11~\text{and}~Alive$}{
    $Default \sim Bern(h(t\mid \mathbf{X},\hat{\boldsymbol{\beta}}))$\;
    \eIf{$Default == 1$}{
        $Alive \gets \text{False}$\;
    }{
        $t \gets t + 1$\;
    }
}
\Return{$t/12$}
\end{algorithm}

\subsection{Fitted Survival Model Coefficients}

\begin{table}[h]
\centering
\begin{tabular}{lrccr}
  \toprule
 & Coef. & \makecell[c]{Robust\\S.E.} & \makecell[c]{Exp(\\Coef.)} & \makecell[c]{p-\\value} \\ 
  \midrule
  Male & 0.408 & 0.021 & 1.503 & \textless.001 \\ 
  Married & -0.239 & 0.024 & 0.787 & \textless.001 \\ 
  APP Channel & 0.170 & 0.034 & 1.186 & \textless.001 \\
  Express Loan & -0.284 & 0.074 & 1.328 & \textless.001 \\
  Repeated Borrower & -0.169 & 0.039 & 0.844 & \textless.001 \\
  Employment: 1 & -0.196 & 0.049 & 0.822 & \textless.001 \\ 
  Employment: 2 & 0.165 & 0.021 & 1.179 & \textless.001 \\ 
  Employment: 3 & 0.027 & 0.028 & 1.027 & .342 \\ 
  Employment: 4 & -0.175 & 0.059 & 0.840 & .003 \\ 
  Education: 1 & -0.513 & 0.056 & 0.599 & \textless.001 \\ 
  Education: 2 & -0.475 & 0.028 & 0.622 & \textless.001 \\ 
  Education: 3 & -0.673 & 0.026 & 0.510 & \textless.001 \\ 
  Education: 4 & -1.383 & 0.278 & 0.251 & \textless.001 \\ 
  ID Province: 1-28 & \multicolumn{4}{c}{\cmark } \\
  \bottomrule
\end{tabular}
\caption{This table reports the main effect of the categorical covariates in the fitted survival model.
The exponentiated coefficient is the multiplicative increase / decrease of hazard compared to the baseline hazard ($h_0(t)$ in Eq.~\ref{eq:generalized_Cox_PH}).}
\label{tab:survival_categorical_coefficients}
\end{table}

\begin{figure}[h] 
\centering
\includegraphics[width=.24\textwidth]{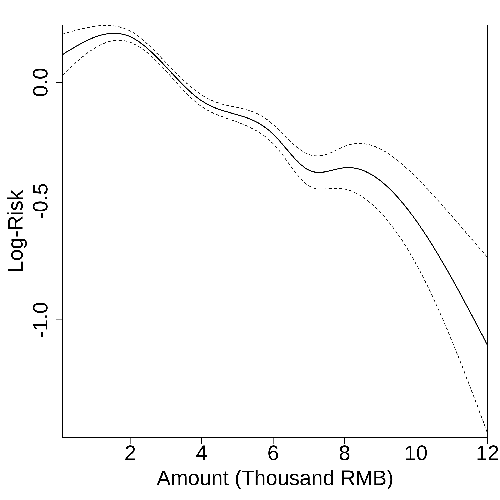}
\includegraphics[width=.24\textwidth]{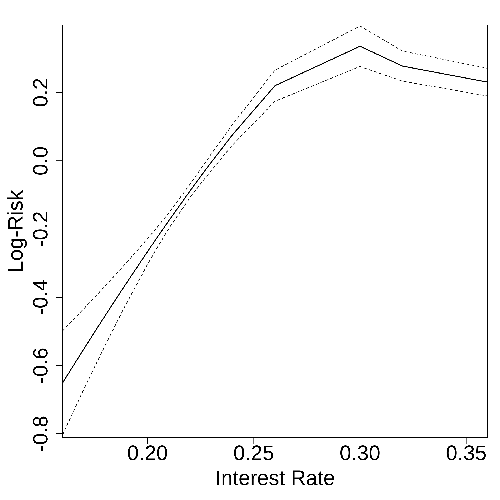}
\includegraphics[width=.24\textwidth]{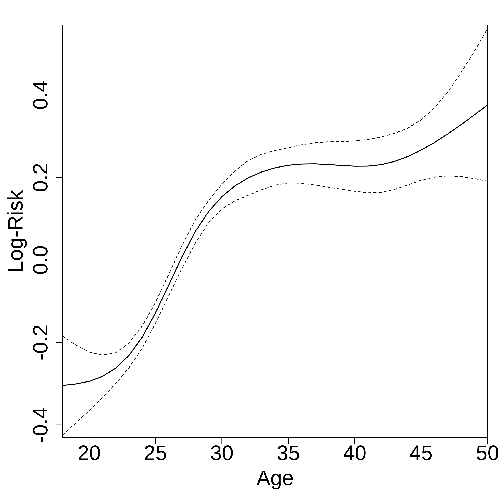} 
\includegraphics[width=.24\textwidth]{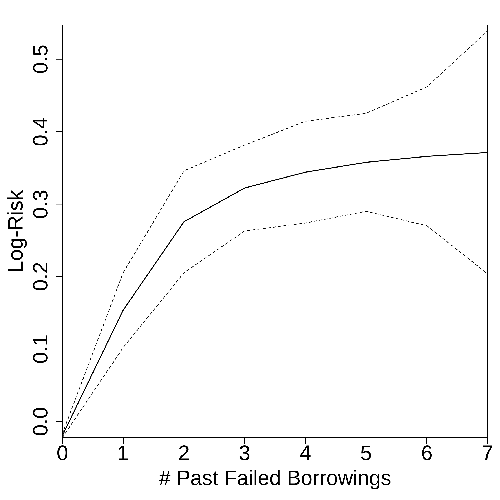} \\
\includegraphics[width=.24\textwidth]{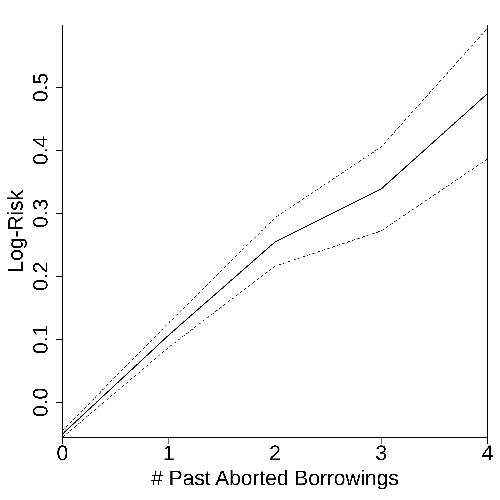}
\includegraphics[width=.24\textwidth]{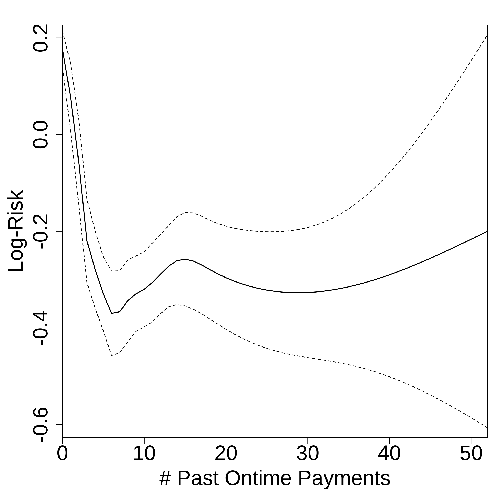} 
\includegraphics[width=.24\textwidth]{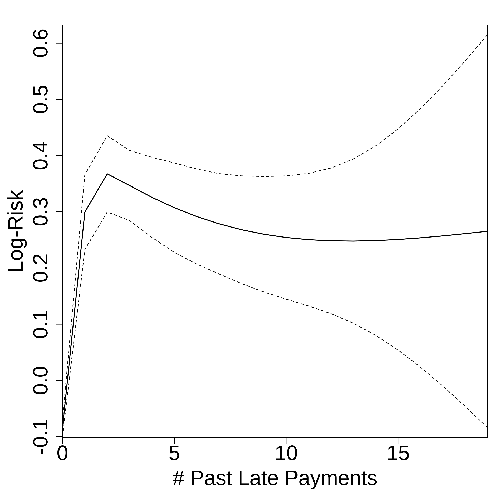} 
\caption{The figures plot the log-risk (with robust S.E.) of the continuous covariates in the fitted survival model.
Exponentiated log-risk is the multiplicative increase / decrease of hazard compared to the baseline hazard ($h_0(t)$ in Eq.~\ref{eq:generalized_Cox_PH}).} \label{fig:survival_term_plot}
\end{figure}

\subsection{Survival Model Diagnostics}
\label{sec:survival_model_diagnostics}
\noindent\textbf{Survival Model is Well-Specified.}
We first assess whether the proportional hazard (PH) assumption is satisfied in our survival model, using the scaled Schoenfeld residual~\cite{schoenfeld1982partial,grambsch1994proportional}.
The Scaled Schoenfeld residual $\epsilon_{i}^{(j)}$ is measured for a specific loan $i$ and a specific covariate, such as age or education, which we index using the superscript $(j)$.
\citet{grambsch1994proportional} shows that, assuming an alternative survival model with time-varing coefficients $\mathbf{\beta}^{(j)}(t)$ for the $j$-th covariate, if $\hat{\mathbf{\beta}}^{(j)}$ is the $j$-th covariate's fitted coefficient from the model of constant coefficients, we have
\begin{equation}
\mathbb{E}[\epsilon_i^{(j)} + \hat{\beta}^{(j)}] \approx \beta^{(j)}(\tau_i),
\end{equation}
where $\tau_i$ is the observation time of the loan $i$.
\textit{I.e.}, the expected value of $\epsilon_i^{(j)} + \hat{\mathbf{\beta}}^{(j)}$ approxiates the actual, potentially time-varying coefficient $\mathbf{\beta}^{(j)}(\tau_i)$.
Using this result, two tests---the $\chi^2$ test and the visual test---can be developed to test for potential violations of the PH assumption.

Tab.~\ref{tab:survival_coxzph_test} reports the $\chi^2$ test, which is a test of zero slop in the regression line fitted between the scaled Schonfeld residual $\epsilon_i^{(j)}$ and observation time $\tau_i$.
For majority of the covaraites, the $\chi^2$ test is unable to reject the null at 0.01 significance level, indicating the PH assumption reasonably holds.
However, for four cavariates ID Province, \# Past Ontime Payments, \# Past Late Payments, and Interest Rate, the $\chi^2$ test significantly rejects the null, indicating violations of the PH assumption.

\begin{figure}[h]
\includegraphics[width=.24\columnwidth]{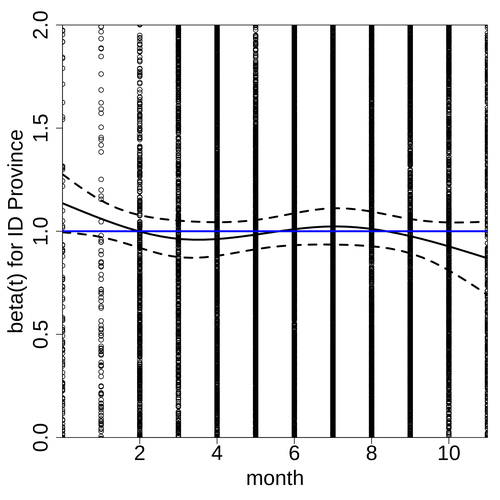} 
\includegraphics[width=.24\columnwidth]{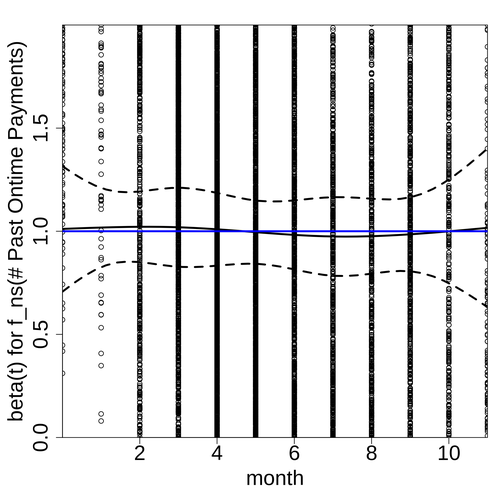}
\includegraphics[width=.24\columnwidth]{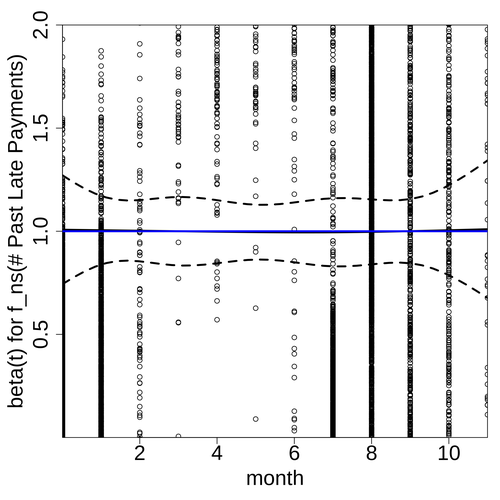}
\includegraphics[width=.24\columnwidth]{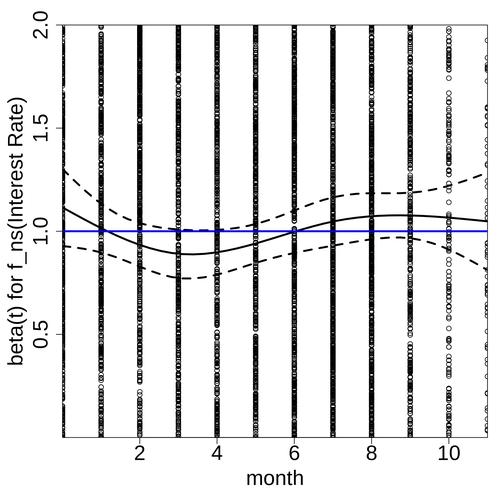}

\centering
\caption{The visual test for PH assumption. A smoothed curve with $95\%$ CI is fitted to the scaled Schoenfeld residual. The blue line is a horizontal fit.} \label{fig:survival_scaled_Schoenfeld}

\end{figure}

To further investigate the degree of these violations, we conduct the visual test for these four covariates, which is shown in Fig.~\ref{fig:survival_scaled_Schoenfeld}.
The visual test fits a smoothed curve to the scaled Schoenfeld residual and compares it against a horizontal fit, which represents a model of constant coefficient.
For these four covariates, the visual test shows the violations are not economically meaningful.
Constant coefficients still provide satisfactory approximations to the smoothed curve, and always lie within the $95\%$ CI.
In fact, although we do not report due to the limited space, the approximation is visually no worse than the other covariates for which the $\chi^2$ test is unable to reject the null.
Combining evidence from both the $\chi^2$ test and the visual test, we show the PH assumption is reasonably satisfied.

\begin{table}[h]
\centering
\begin{tabular}{lrcr}
\toprule
&\multicolumn{1}{c}{$\chi^{2}$}&\multicolumn{1}{c}{df}&\multicolumn{1}{c}{p-value} \\
\midrule
  Male & 0.521 & 1 & 0.470 \\ 
  Married & 5.268 & 1 & 0.022 \\ 
  APP Channel & 3.630 & 1 & 0.057 \\ 
  Express Loan & 5.466 & 1 & 0.019 \\ 
  Repeated Borrower & 5.626 & 1 & 0.018 \\ 
  Employment & 2.286 & 4 & 0.683 \\ 
  Education & 1.698 & 4 & 0.791 \\ 
  ID Province & 76.711 & 28 & \textless0.001 \\ 
  $f_{ns}$(Age) & 11.596 & 6 & 0.072 \\ 
  $f_{ns}$(\# Past Failed Borrowings) & 12.942 & 4 & 0.012 \\ 
  $f_{ns}$(\# Past Aborted Borrowings) & 3.746 & 4 & 0.442 \\ 
  $f_{ns}$(\# Past Ontime Payments) & 59.121 & 9 & \textless0.001 \\ 
  $f_{ns}$(\# Past Late Payments) & 32.308 & 5 & \textless0.001 \\ 
  $f_{ns}$(Amount) & 1.903 & 6 & 0.928 \\ 
  $f_{ns}$(Interest Rate) & 136.280 & 4 & \textless0.001 \\ 
\bottomrule
\end{tabular}
\caption{The $\chi^2$ test for PH assumption. The null hypothesis is the regression line fitted between the scaled Schoenfeld residual and the observation time has zero slope.}  \label{tab:survival_coxzph_test}
\end{table}

\begin{figure}[h!]
\begin{minipage}{0.48\columnwidth}
\centering
\includegraphics[width=0.9\columnwidth]{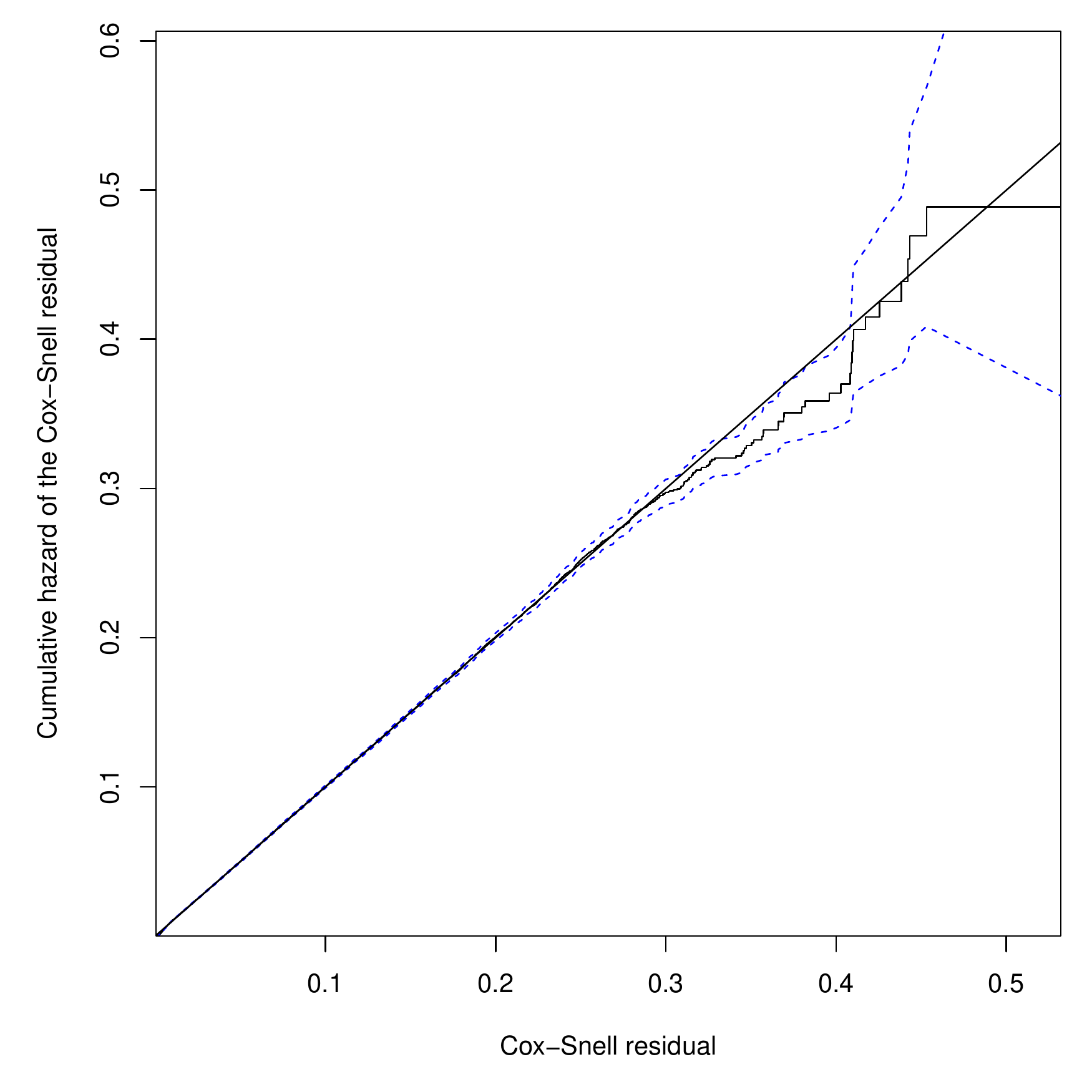}
\caption{Cox-Snell residual with 95\% CI, compared to the 45\degree~line.} \label{fig:survival_cox_snell}
\end{minipage}
\hfill
\begin{minipage}{0.48\columnwidth}
\centering
\includegraphics[width=0.9\columnwidth]{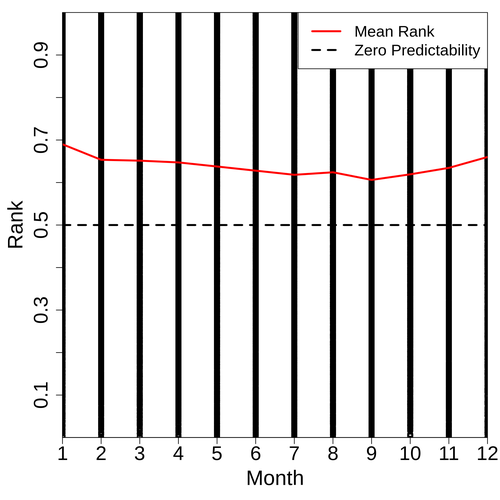}
\caption{Defaulted loans' rank as predicted by the survival model.} \label{fig:survival_rank}
\end{minipage}
\end{figure}

\noindent\textbf{Survival Model Has Goodness-of-Fit.}
We use the Cox-Snell residual~\cite{cox1968general} to assess the model's goodness-of-fit.
If the model is correct and well-fitted, the Cox-Snell residual $\epsilon_{CS}$ should resemble a censored sample from an unit exponential distribution.
And if $\epsilon_{CS}$ follows a censored unit exponential, the cumulative hazard of $\epsilon_{CS}$ against $\epsilon_{CS}$ should be a straight line with zero intercept and unit slope.
Fig.~\ref{fig:survival_cox_snell} implements this test. 
The estimated cumulative hazard of $\epsilon_{CS}$ closely matches the 45\degree~line and thus the survival model is well-fitted.

\noindent\textbf{Survival Model is Predictive.}
The fitted survival model is predictive of the borrower's default with a concordance index 0.638 (S.E.=0.001)~\cite{harrell1996multivariable}.
It means the fitted model predicts which loan has higher repayment ratio---or is more trustworthy---with $63.8\%$ accuracy, for all possible pairs of comparable loans.
In Fig.~\ref{fig:survival_rank}, we plot the defaulted loans' ranks as predicted by the survival model against their default months.
Each defaulted loan's rank ranges from 0 to 1.
A rank of 1 means the loan has the highest predicted default probability at its default month from that month's risk set, which includes all loans that have not defaulted so far.
A rank of 0 then means the loan has the lowest predicted default probability.
We see that the survival model correctly predicts higher risk for the actually defaulted loans across all motnhs.
Its predictability has mild variation and is the highest at the very first month.

\newpage
\subsection{Analysis of Disparate Impact}

\begin{table}[h!]
\centering
{\small
\begin{tabular}{l | lccc}
  \toprule
 &  & Estimate & \makecell{Lower\\$95\%$ CI} & \makecell{Upper\\$95\%$ CI} \\
  \midrule
  \multirow{3}{*}{Month}
  & Jan. -- Mar. & $-0.0350$ & $-0.0354$ & $-0.0348$ \\
  & Apr. -- Jun. & $-0.0374$ & $-0.0376$ & $-0.0372$ \\
  \midrule
  \multirow{2}{*}{Married} & No & $-0.0458$ & $-0.0463$ & $-0.0457$ \\
  & Yes & $-0.0321$ & $-0.0322$ & $-0.0318$ \\
  \midrule
  \multirow{2}{*}{\makecell[l]{Repeated\\Borrower}} & No & $-0.0219$ & $-0.0220$ & $-0.0217$ \\
  & Yes & $-0.0306$ & $-0.0310$ & $-0.0303$ \\
  \midrule
  \multirow{2}{*}{Age} & $<26$ & $-0.0517$ & $-0.0521$ & $-0.0516$ \\
  & $\geq 26$ & $-0.0338$ & $-0.0339$ & $-0.0335$ \\
  \midrule
  \multirow{2}{*}{\makecell[l]{Amount\\(k RMB)}} & $<2.5 $ & $-0.0355$ & $-0.0360$ & $-0.0352$ \\
  & $\geq 2.5$ & $-0.0390$ & $-0.0392$ & $-0.0389$ \\
  \midrule
  \multirow{2}{*}{Interest Rate} & $\leq0.22$ & $-0.0455$ & $-0.0458$ & $-0.0454$ \\
  & $>0.22$ & $-0.0332$ & $-0.0324$ & $-0.0319$ \\
  \midrule
  \multirow{2}{*}{Employment} & Unknown & $-0.0241$ & $-0.0244$ & $-0.0238$ \\
  & Student & $-0.0645$ & $-0.0662$ & $-0.0644$ \\
  & Worker & $-0.0437$ & $-0.0440$ & $-0.0435$ \\
  & Self-employed & $-0.0332$ & $-0.0334$ & $-0.0328$\\
  & Online Shop Owner & $-0.0551$ & $-0.0574$ & $-0.0524$\\
  \midrule
  \multirow{2}{*}{Education} & Unknown & $-0.0438$ & $-0.0441$ & $-0.0437$ \\
  & Junior College & $-0.0453$ & $-0.0488$ & $-0.0443$ \\
  & College & $-0.0315$ & $-0.0320$ & $-0.0312$ \\
  & Bachelor & $-0.0209$ & $-0.0217$ & $-0.0204$ \\
  & Master or Doctorate & $0.0225$ & $0.0120$ & $0.0284$ \\

  \bottomrule
\end{tabular}
}
\caption{Average DI in Disaggregated Data Subsets.}
\label{tab:UGD_disaggregated}
\vspace{-2ex}
\end{table}

\subsection{Bayesian Inference}
\label{sec:bayesian_inference}

The decision model explained in Sec.~\ref{sec:decision_model} can be written as follows:
\begin{align*} \label{eq:one_sided_final_model_P2P}
D_i \mid (G_i=g) \sim &
Bern ( p=\Phi (  \frac{1}{\sigma_{g,1}} \lambda_i - \frac{1}{\sigma_{g,1}} \times(\pi_g / \gamma_g) \times \frac{1}{1+R_i}   + \frac{1}{\sigma_{g,1}} \times (1 / \gamma_g - 1) \mu_g ) ), \\
\gamma_g =& \frac{(\sigma_{g,1})^{-2}}{(\sigma_{g,0})^{-2}+(\sigma_{g,1})^{-2}}.
\end{align*}

This model can be directly converted to a Bayesian latent variable model with the following unknown parameters: the repayment ratio noise variance $\sigma_{f,1}, \sigma_{m,1}$ and the decision threshold $\pi_f, \pi_m$.
To complete the Bayesian model specification, we put the following weakly informative priors:
\begin{align*}
1/\sigma_{f,1}, 1/\sigma_{m,1} &\sim N^+ (0, 2), \\
\pi_m, \pi_f &\sim N^+(0, 2).
\end{align*}
We use $N^+$ to denote the half-Normal distribution.
We put priors on the the inverse of $\sigma_{f,1}$ and $\sigma_{m,1}$ because we find it stabalizes the sampling procedure.

Performing Bayesian inference directly with this model, however, is compute-intense because the number of loans is more than half a million.
We leverage the fact that both interest rate and repayment ratio only take some discrete values, and reframe the model as a Binomial variable at unique combinations of interest rate and repayment ratio, similar to~\citet{simoiu2017problem}.
After reformulation, the number of data is reduced to less than 500.
The reformulation from a Bernoulli random variable to a Binomial random variable is without any loss and dramatically speeds-up inference.

We estimate the posterior distribution of $\{\sigma_{g,1},\pi_{g}\}_{g\in\{m,f\}}$ using Hamiltonian Monte Carlo (HMC)~\cite{duane1987hybrid,neal2011handbook}, a kind of Markov chain Monte Carlo (MCMC) sampling approach~\cite{metropolis1953equation,hastings1970monte}.
Specifically, we use the No-U-Turn sampler (NUTS)~\cite{hoffman2014no} implemented in the \texttt{PyMC3} package~\cite{salvatier2016probabilistic} to sample from the posterior distribution.
We set the target acceptance rate to $0.98$, use 5000 draws for both warmup and estimation, and run 4 chains in parallel.

To yield confidence interval estimates, we similarly bootstrap both stages of 2SPS.
The difference here is the second stage returns an estimated posterior distribution rather than a point estimate.
We combine posteriors from different iterations by uniform averaging.
Averaging is a common technique to estimate the true ``consensus'' posterior from a set of subposteriors~\cite{huang2005sampling,angelino2016patterns,scott2016bayes}.
It can be shown bootstrapping the 2SPS procedure is in fact approximating Bayesian inference on a richer hierarchical Bayesian model that acknowledges the uncertainty of the predicted repayment ratios.

\subsection{Visual Inspection of the Traces}

\begin{figure}[h!] 
\begin{subfigure}[h]{.8\textwidth}
\includegraphics[width=\textwidth]{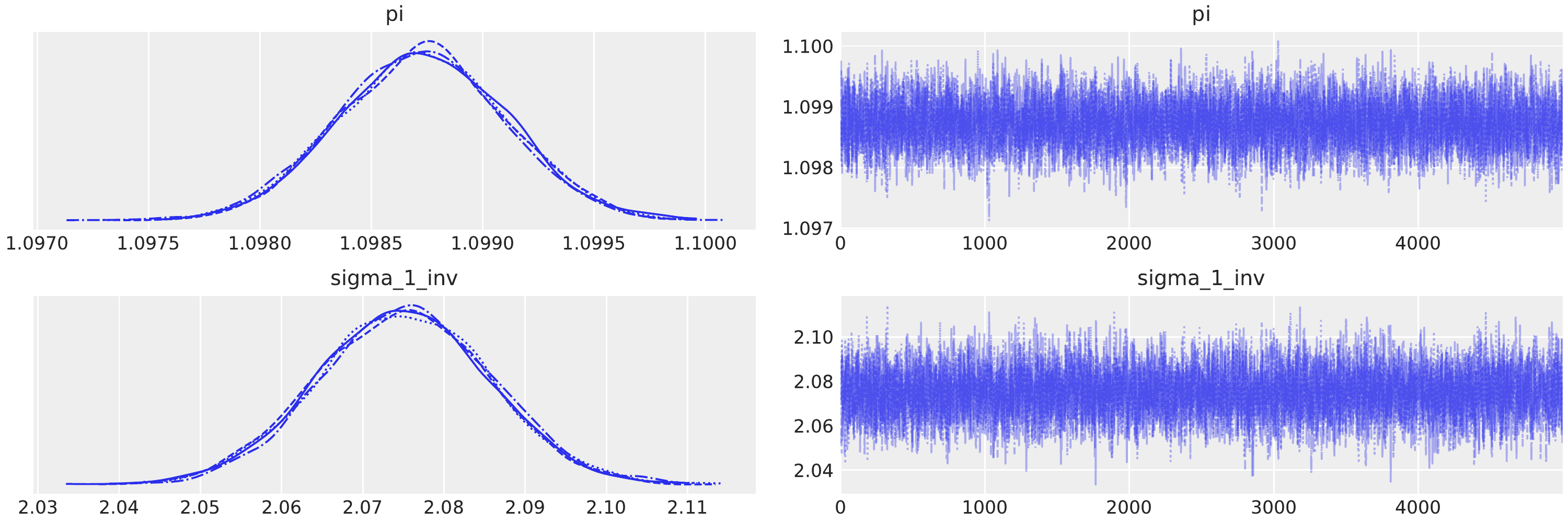}
\caption{Female Borrowers.}
\end{subfigure}
\\
\begin{subfigure}[h]{.8\columnwidth}
\includegraphics[width=\textwidth]{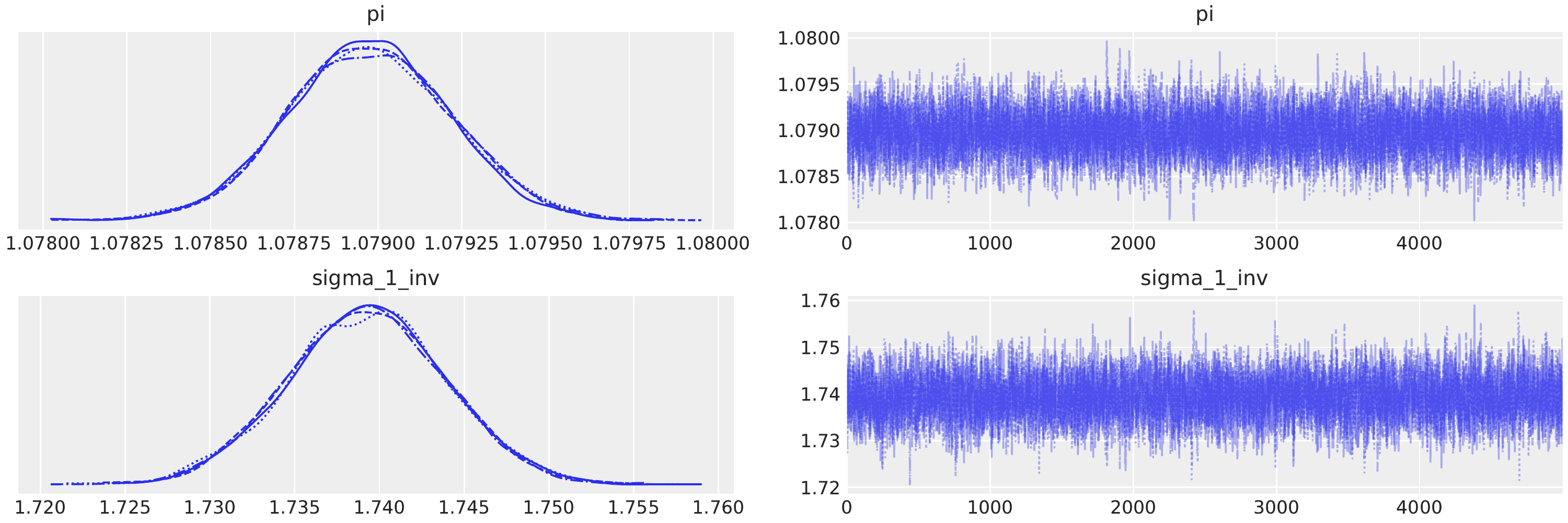}
\caption{Male Borrowers.}
\end{subfigure}
\caption{Trace Plot From the Bayesian Inference Step. pi denotes the funding threshold $\pi_g$ and sigma\_1\_inv denotes $1/\sigma_{g,1}$.} \label{fig:Bayesian_traces}
\end{figure}

\end{document}